\documentclass[12pt,preprint]{aastex}
\received{2003 August 16}
\begin{document}

\title{Gamma-ray Emission Properties from Mature Pulsars in the Galaxy and in the Gould Belt}
\author{K.S. Cheng$^1$, L. Zhang$^{2,3}$, P. Leung$^1$, Z.J. Jiang$^3$}
\affil{$^1$Department of Physics, the University of Hong Kong,
Hong Kong, PRC\\
$^2$National Astronomical Observatories/Yunnan Observatory,Chinese Academy of Sciences, P.O. Box 110, Kunming, PRC\\
$^3$ Department of Physics, Yunnan University, Kunming, PRC }

\begin{abstract}
We study the $\gamma$-ray emission properties of pulsars by using
a new self-consistent outer gap model. The outer gap can exist in
pulsars with age over million years old if the effect of magnetic
inclination angle as well as the average properties of the outer
gap are considered. The mature $\gamma$-ray pulsars, whose ages
are between 0.3 to 3 million years old,  are able to move up to
high galactic latitude. Moreover, their $\gamma$-ray luminosity
are weaker and their spectra are softer than those of younger
$\gamma$-ray pulsars in the galactic plane significantly. We use a
Monte Carlo method to simulate the statistical properties of
$\gamma$-ray pulsars in the Galaxy as well as in the Gould Belt.
We find that $\gamma$-ray pulsars located at $\mid b \mid <
5^{\circ}$ and located at $\mid b \mid > 5^{\circ}$ have very
different properties. High galactic latitude $\gamma$-ray pulsars
are dominated by mature pulsars with longer periods, weaker fluxes
and softer spectra. If the pulsar birth rate in the Galaxy and the
Gould Belt are $\sim 10^{-2}yr^{-1}$ and $\sim 2\times
10^{-5}yr^{-1}$ respectively, there are 42 and 35 radio-quiet
$\gamma$-ray pulsars for $\mid b \mid < 5^{\circ}$ and $\mid b
\mid > 5^{\circ}$ respectively. Radio-quiet $\gamma$-ray pulsars
from the Gould Belt are 2 and 13 for $\mid b \mid < 5^{\circ}$ and
$\mid b \mid > 5^{\circ}$ respectively. We suggest that a good
fraction of unidentified EGRET $\gamma$-ray sources may be these
radio-quiet $\gamma$-ray pulsars. Furthermore $\gamma$-ray pulsars
located at $\mid b \mid > 5^{\circ}$ satisfies $L_{\gamma} \propto
L^{\beta}_{sd}$ whereas $L_{\gamma} \propto L^{\delta}_{sd}$ for
$\gamma$-ray pulsars in the galactic plane, where $\beta \sim 0.6$
and $\delta \sim 0.3$ respectively.

\end{abstract}


\section{Introduction}

There are 170 unidentified $\gamma$-ray sources in the third EGRET
catalog (Hartman et al. 1999), where $\sim$50 sources close to the
Galactic plane with $|b|<5^{\circ}$ and $\sim$70 sources in the
medium latitudes with $|b|$ between $5^{\circ}$ and $30^{\circ}$.
For those unidentified $\gamma$-ray sources in the Galactic plane,
many of them are associated with Wolf-Rayet and Of stars, SNRs and
OB stars(Montmerle 1979; Kaaret \& Cottam 1996; Yadigaroglu \&
Romani 1997, Romero et al. 1999). Most of these objects are
considered as pulsar tracers, therefore it is natural to suggest
that these low latitude sources may be Geminga-like pulsars, which
are radio-quiet pulsars (Yadigaroglu \& Romani 1995; Cheng \&
Zhang 1998; Zhang, Zhang \& Cheng 2000). Since $\gamma$-ray
pulsars are known to be steady $\gamma$-ray emitters, it has been
suggested that the variability should be a good indicator to
identify the real pulsar candidates from the unidentified EGRET
sources (Romero, Combi \& Colomb 1994; Mclaughlin et al. 1996;
Zhang et al. 2000; Torres et al. 2001; Nolan et al. 2003). Gehrels
et al. (2000) define a class of sources  as steady: a source is
steady if the most significant detection of a  source in 3EG
catalogue is for a timescale of years and if that particular flux
is within $3\sigma$ of the flux calculated for the full data set.
 From this classification, 48 unidentified sources at $|b|<
5^{\circ}$  and 72 unidentified sources at $|b|> 5^{\circ}$ are
steady sources respectively. On the other hand, Grenier (2001)
suggests another definition called persistent sources, which are
those detected sources with a significance $\sqrt{TS}>4$ for every
observation at $|b|>2.5^{\circ}$. If we assume that the sources
with a significance $\sqrt{TS}>5$ at $|b|\le 2.5^{\circ}$ are
persistent sources, there are 40 (45) persistent unidentified
EGRET sources at $|b|< 5^{\circ}$ ($|b|> 5^{\circ}$).

However, the spectral properties of  medium latitude sources are
significantly softer, fainter and have a steeper logN-logS
function than those at low latitudes (Gehrels et al. 2000). It has
been suggested that they are associated with recent supernovae in
the nearby Gould Belt (Grenier 1997; Gehrels et al. 2000; Grenier
2000). Their natures remain as mystery. Harding and Zhang (2001)
used the polar cap models (Daugherty \& Harding 1996; Harding \&
Muslimov 1998) to investigate whether $\gamma$-ray pulsars viewed
at a large angle to the neutron star magnetic pole could
contribute to unidentified EGRET sources in the medium latitudes
associated with the Gould Belt. They suggest that the off-beam
$\gamma$-rays come from high-altitude curvature emission of
primary particles can radiate over a large solid angle and have a
much softer spectrum than that of the main beams, and at least
some of radio-quiet Gould Belt sources detected by EGRET could be
such off-beam $\gamma$-ray pulsars.

Recently, the brightest of unidentified EGRET sources in the
medium latitude, 3EG J1835+5918, is strongly suggested as the
second Geminga-like pulsar ( radio-quiet pulsar) by
multi-wavelength observations including Chandra, HST and Jodrell
Bank (Mirabal et al. 2000; Mirabal \& Halpern 2001; Halpern et al.
2002). The X-ray spectrum can be described by two components: a
soft thermal X-ray spectrum with a characteristic temperature
($T_{\inf} \approx 3\times 10^5$K and a power law hard tail with a
photon index $\gamma \approx $2, which closely resembles to the
X-ray spectrum of Geminga. The repeated radio observations at
Jodrell Bank did not show any periodicity. The X-ray data suggests
that the distance is between 250-800pc, which is consistent with
the distance to the Gould Belt.

In this paper, we use the revised outer gap models (Zhang et al.
2004) to investigate emission properties of $\gamma$-ray pulsars
in the Galaxy as well as in the Gould Belt. This revised outer gap
model is based on the original outer gap models (Cheng, Ho \&
Ruderman 1986a, {\bf{hereafter CHR I}}, 1986b; Zhang \& Cheng
1997) but takes into account the effect of the inclination angle
($\alpha$), which is the angle between the magnetic axis and the
rotation axis, in determining the size of the outer gap,which is
defined as the ratio between the dimension of the gap
perpendicular to the magnetic field and the light cylinder radius.
The revised model also takes into account the fact that the
typical radiation region of the outer gap is not necessary at half
of the light cylinder instead it should be better represented by
an appropriate average over the entire outer gap. This effect is
particular important for old pulsars. If the outer gap is only
represented at half of the light cylinder, then the outer gap is
assumed to be turn off when the gap size at this region is larger
than unity.  This new model has taken the entire active region of
the outer gap into account. As long as the gap size is less than
unity in some parts of magnetosphere, the outer gap still exist.
This effect allows some pulsars with appropriate combination of
$\alpha$, $P$ and $B$, can maintain their outer gaps until a few
million years old. These pulsars are able to move up to high
galactic latitude and their ages make them weak $\gamma$-ray
sources. Furthermore, we will show that these pulsars will emit
softer spectra and their relation between $\gamma$-ray luminosity
and spin-down power differs from that of younger pulsars located
in the galactic plane. We organize the paper as follows. In
section 2, we review the new outer gap models. In section 3, we
study the $\gamma$-ray emission properties of pulsars by using the
new self-consistent outer gap model. In section 4, we describe a
Monte Carlo simulation method to determine the statistical
distributions of radio-quiet and radio-loud $\gamma$-ray pulsars.
In section 5, we summarize the simulation results and discuss
their implications. Finally, a brief conclusion is given in
section 6.

\section{Outer Gap Models}

Zhang \& Cheng (1997) have proposed a self-consistent mechanism to
describe the high-energy radiation from the rotation-powered
pulsars. In this model, the radiation mechanism of relativistic
charged particles from a thick outermagnetospheric accelerator
(outer gap) is synchro-curvature radiation (Cheng \& Zhang 1996)
and the characteristic energy of high-energy photon emitted from
the outer gap is determined by the pulsar global parameters, i.e.
rotation period P and the dipolar magnetic field B, as well as the
fractional size of the outer gap ($f$), which is a ratio between
the mean vertical separation of the outer gap boundaries in the
plane of the rotation axis and the magnetic axis to the light
cylinder radius, and is given by
\begin{equation}
E_{\gamma}(f)\approx 5.0 \times
10^7f^{3/2}B_{12}^{3/4}P^{-7/4}\left(\frac{r}{R_L}\right)^{-13/8}
eV
\end{equation}
where $B_{12}$ is the dipolar magnetic field in unites of
$10^{12}$G, $R_L$ is the light cylinder radius and $r$ is the
distance to the neutron star. The $\gamma$-ray spectrum becomes
exponentially decay beyond $E_{\gamma}(f)$. The fractional size of
the outer gap determines the total $\gamma$-ray luminosity from
the outer gap (Zhang \& Cheng 1997) and is given by
\begin{equation}
L_{\gamma} = f^3L_{sd}
\end{equation}
where $L_{sd} = 3.8 \times 10^{31} B_{12}^2 P^{-4} erg s^{-1}$ is
the spin-down power of pulsar. CHRI has shown that the inner
boundary of the static outer gap begins at the null charge
surface(${\bf{\Omega}}\cdot {\bf{B}}=0$). Recently Hirotani \&
Shibata (2001) has shown that if there is injected current from
the inner boundary, the position of the inner boundary can shift
either toward the star or even close to the light cylinder
depending on the sign and the magnitude of the injected current.
However, it is not clear what causes this injected current. In
this paper we shall assume that the outer gap begins at the null
charge surface. The radial distance from the null surface surface
to the star is $r_{in}$, which is a function of the magnetic
inclination angle $\alpha$ and is given by ${r_{in}\over
R_L}={\sin^2(\theta_{in}-\alpha)\over \sin\theta_c
\sin^2(\theta_c-\alpha)}$, where $\theta_{in}$ is the polar angle
between of $r_{in}$ determined by $\tan\theta_{in}={1\over
2}(3\tan\alpha+\sqrt{9\tan^2\alpha+8})$ and $\theta_c$ is the
polar angle to the position where the first open field line
intercepts the light cylinder determined by $\tan\theta_c=-{3\over
4\tan\alpha}(1+(1+8\tan^2\alpha/9)^{1/2})$. The fractional size of
the outer gap is limited by the pair production between the soft
thermal X-rays with characteristic energy $E_x$ from the stellar
surface and the high-energy photons with energy $E_{\gamma}(f_o)$
emitted from the outer gap. The energy of the soft X-ray photons
is determined by the backflow of the primary electrons/positrons.
Each of these backflow particles can still maintain about
10.6P$^{1/3}$ ergs and deposit on the stellar surface. This energy
will be emitted as soft thermal X-rays from the stellar surface
(Halpern \& Ruderman 1993), whose characteristic energy is given
by $E_x(f_o)\approx 1.2 \times 10^2
f_o^{1/4}B_{12}^{1/4}P^{-5/12}$eV. Although the thermal X-ray
photon density is low, every pair resulting from X-ray and
high-energy photon interactions can emit $\sim 10^5$ high-energy
photons when they are accelerated in the gap. Such a large
multiplicity can produce sufficient number of $e^{\pm}$ pairs as
to sustain the outer gap. Assuming $r=R_L/2$ and head-on
collision, they obtained the fractional size of the outer gap as
\begin{equation}
f_o(B,P) \approx 5.5P^{26/21}B_{12}^{-4/7}
\end{equation}
from the condition of photon-photon pair production
$E_xE_{\gamma}(1-\cos(\theta_{X\gamma}))=2(m_ec^2)^2$.

However, this model did not take into account the fact that when
the magnetic inclination angle ($\alpha$) becomes large, the
characteristic energy of high-energy photons from the gap, which
depends on $\alpha$, also increases. In fact both $E_x$ and
$\theta_{X\gamma}$ also depend on $\alpha$. Most important, Zhang
and Cheng (1997) has assumed a typical distance
($r=\frac{R_L}{2}$) to represent the outer gap. In other words,
most radiation and pair production activities are assumed to take
place at this characteristic region. They assume that when
$f_o(r=\frac{R_L}{2}, B,P) > 1.$ the outer gap does not exist.
However, even the mid-distance to the outer gap depends on the
inclination angle because the outer gap begins at the null charge
surface, which depends on $\alpha$. Realistically, as long as
$e^{\pm}$ pairs can be produced beyond the null surface, the outer
gap should be still active, namely, accelerating charged
particles. Of course, for regions with $f(r)> 1.$ the gap
potential should drop off rapidly and hence becomes unimportant.
Therefore the active region of the outer gap should begin at null
surface and stop at a distance $r_b$ where $f(r_b)= 1.$ (Zhang et
al. 2004). In explaining the detail $\gamma$-ray spectrum of a
given pulsar, it is important to know the radiation coming from
which parts of the pulsar magnetosphere (Cheng, Ruderman \& Zhang
2000). However, in order to study the statistical properties of
pulsars a representative region is a very useful approximation.
They assume that the representative region of the outer gap is the
averaging distance to the gap, they obtain the mean fractional
size of the gap $f(P,B,\alpha)$, which can be approximately
expressed as
\begin{equation}
f(\alpha,B,P) \approx f_o(B,P)\eta(\alpha,B,P)
\end{equation}
where $\eta(\alpha,B,P)$ is a monotonically function of $\alpha$,
$B$ and $P$.  It roughly decreases by a factor 3 from young
pulsars with large inclination angle to old pulsars with small
inclination angle. The integral expression of $f(\alpha,B,P)$ is
given by Eq. (38) and the variation of $\eta$ for different
pulsars is given in figure 3 of Zhang et al. (2004) respectively.
Although the variation of $\eta$ is only a factor of 3, the
implications are very important. First the cut-off period of
$\gamma$-ray pulsars is about a factor of $3^{21/26} = 2.4$ longer
for fixed $\alpha$ and $B$ in comparing with the old model.
Secondly the cut-off age of $\gamma$-ray pulsars is about a factor
of 5.8 longer as well. This means that there are a lot more
$\gamma$-ray pulsars in high galactic latitude than previously
expected. These $\gamma$-ray pulsars with older age can move up to
the high galactic latitude and may contribute to the unidentified
EGRET sources.

\section{Gamma-ray Emission Properties of Mature Pulsars}

For the thick outer gap (Zhang \& Cheng 1997), $\gamma$-rays are
produced inside the thick outer gap by curvature radiation from
the primary $e^{\pm}$ pairs along the curved magnetic field lines.
However, Cheng \& Zhang (1996) studied the radiation from the
charged particles in the curved magnetic field, and pointed out
that the radiation should be described more accurately by a
general radiation mechanism called synchro-curvature radiation
mechanism, in which the radiation is being emitted by the charged
particles moving in a spiral trajectory along the curved magnetic
field lines. This mechanism differs from synchrotron and curvature
mechanisms in general, but reduces to either synchrotron radiation
when the radius of curvature of the local magnetic field lines is
infinite or to curvature radiation when the pitch angle is zero.
In fact, when the synchrotron gyro-radius $r_B=\gamma
mc^2\sin\theta_p/eB(r)$ and the curvature radius of field
$s\approx \sqrt{rR_L}$ is comparable, the synchro-curvature
mechanism really provides a significant improvement, where
$\gamma$ is the Lorentz factor of the accelerated particles and
$\theta_p$ is the pitch angle of the charged particles in the
curved magnetic field. Zhang \& Cheng (1997) used this mechanism
to describe the production of non-thermal photons from the primary
$e^{\pm}$ pairs along the curved magnetic field lines in the outer
gap. The primary $e^{\pm}$ pairs have an approximate power-law
distribution inside the outer gap because the energy and density
of the primary $e^{\pm}$ pairs depend on local values of magnetic
field, electric field and radius of curvature. In steady state,
the energy distribution of the accelerated particles in the outer
gap is $(dN/dE_e)\propto E^{-16/3}_e$, where $E_e$ is the energy
of the accelerated particle. Using the general formula of the
synchro-curvature radiation power spectrum given by Cheng \& Zhang
(1996) and $(dN/dE_e)dE_e=(dN/dx)dx$, where $x=s/rR_L$, the
differential flux at the Earth is (Zhang \& Cheng 1997)
\begin{equation}
F(E_{\gamma})\approx {1\over \Delta\Omega d^2} {\dot{N}_0\over
E_{\gamma}} \int^{x_{max}}_{x_{min}}x^{3/2}{R_L\over R_c}
\left[\left(1+{1\over R^2_cQ^2_2}\right)F(y)- \left(1-{1\over
R^2_cQ^2_2}\right)yK_{2/3}(y)\right]dx \;\;, \label{flux}
\end{equation}
where $\Delta\Omega$ is the solid angle of $\gamma$-ray beaming,
$d$ is the distance to the pulsar,
$\dot{N}_0=\sqrt{3}e^2\gamma_0N_0/ hR_L$, $N_0\approx
1.4\times10^{30}f(B_{12}/P)R_6^3$, $\gamma_0\approx 2\times
10^7f^{1/2}(B_{12}/P)^{1/4}$,
$R_c=xR_L/[(1+r_B/(xR_L))\cos^2\theta_p
+(R_L/r_B)x\sin^2\theta_p]$,
 $Q_2=(1/xR_L)[((r_B/xR_L)+1-3(R_L/r_B)x)
\cos^4\theta_p+3(R_L/r_B)x\cos^2\theta_p+(R_L/r_B)^2x^2$
$\sin^4\theta_p]^{1/2}$,
 and  $\sin\theta_p\approx 0.79f^{1/2}B^{-3/4}_{12}P^{7/4}x^{17/4}$;
 $F(y)=\int^{\infty}_{y}K_{5/3}(z)dz$, where $K_{5/3}$
is the modified Bessel functions of order 5/3, $y=E_{\gamma}/E_c$
and $E_c=(3/2)(\hbar c\gamma^3/x)(xQ_2)$ is the characteristic
energy of the synchro-curvature photons. $x_{min}$ and $x_{max}$
are the minimum and maximum values of $x$. For $x_{min}$, it can
estimated as $x_{min}=(r_{in}/R_L)^{1/2}$. The maximum value of
$x$ can be estimated by $x_{max} \approx
\frac{\theta_p}{\theta_f}\sqrt{r_f/R_L}$ (Arons \& Scharlemann
1979), where $\theta_p$ is the angular width of the polar cap,
$\theta_f$ is the angular width between the magnetic axis to the
upper boundary field lines of the outer gap and $r_f$ is the
radial distance to the upper boundary field lines intercepting
with the light cylinder. Zhang \& Cheng (1997) has discussed the
possible value of $x_{max}$. In general the $\gamma$-ray beaming
solid angle should be different for various $\gamma$-ray pulsars,
which is a function of the magnetic inclination angle as well as
the size of the outer gap(Zhang et al. 2004). Some approximate
forms of beaming solid angle have been given(Yadigaroglu \& Romani
1995; Zhang, Zhang \& Cheng 2000). How accurate of these
approximation forms are not known. For simplicity, in this paper,
we will treat these two parameters, $\Delta\Omega$ and $x_{max}$
as constants. By fitting the $\gamma$-ray spectrum of known
$\gamma$-ray pulsars, these two parameters are chosen to be
$\Delta\Omega \sim 1 sr$ and $x_{max} \sim 2$ respectively (Cheng
\& Zhang 1998).

In order to compare with the EGRET data, the integral flux is
necessary and is given by
\begin{equation}
F(\ge 100MeV)=\int^{E_{max}}_{100MeV}F(E_{\gamma})dE_{\gamma}
\end{equation}
where $E_{max}$ is the maximum energy of gamma-rays and is chosen
to be 100GeV.

In Fig. 1 - Fig. 3,  the model $\gamma$-ray spectra are calculated
for various periods, magnetic fields and inclination angles
respectively. We can see that the spectra become softer when the
period decreases, the magnetic field increases or the inclination
angle decreases in the energy range of EGRET (100MeV-10GeV).
Alternatively speaking the spectral break becomes larger when the
period increases, the magnetic field decreases or the inclination
angle increases. We can understand these trends by examining the
behavior of the energy break at Eq. 1. Typically most power
radiated from the primary electrons/positrons come from regions
with characteristic distance of order of $\sim R_L$. Substituting
Eq. 3 and Eq. 4 into Eq. 1, we obtain the spectral break is
$\propto \eta^{3/2} P^{3/28} B^{-3/28}_{12}$, which explains the
relations between spectral variations in Fig. 1-3. However, in
subsequent sections we will find that the key differences between
$\gamma$-ray pulsars in the galactic plane and those in the high
galactic latitude are: (1)galactic plane $\gamma$-ray pulsars are
younger, shorter in period and have larger inclination angles and
(2)they satisfy different relation between $L_{\gamma}$ and
$L_{sd}$. In fact, most high latitude $\gamma$-ray pulsars satisfy
$L_{\gamma}\propto L_{sd}$ which means $f \sim 1$ (cf. Fig. 6f and
the lower panel of Fig. 7). In this case, the spectral break is
$\propto \eta^{21/16} P^{-1/8}$. Since galactic $\gamma$-ray
pulsars have larger inclination angles and shorter periods in
comparing with the high galactic latitude $\gamma$-ray pulsars, so
the spectrum of $\gamma$-ray pulsars in high galactic latitude has
a softer spectrum than those in the galactic plane. In Fig. 4, we
compare the typical spectra between the high and low galactic
latitude $\gamma$-ray pulsars.

\section{Monte Carlo Simulation of $\gamma$-ray Pulsars in the Galaxy and in the Gould Belt}

In order to consider the $\gamma$-ray luminosity and spatial
evolution of pulsars in the Galaxy and in Gould Belt, the initial
values of parameters of pulsar at birth, which include the initial
position, velocity, period and magnetic field strength, are
needed. The procedure of Monte Carlo method and the evolution of
pulsar parameters are described in our previous works (Cheng \&
Zhang 1998; Zhang \& Cheng 1999; Zhang, Zhang \& Cheng 2000; Fan,
Cheng \& Manchester 2001; Zhang et al. 2004). Here, we briefly
describe basic assumptions for generating the $\gamma$-ray pulsars
(radio-quiet and radio-loud) in the Galaxy as well as in the Gould
Belt:

\begin{description}
\item{1} The pulsars are born at a rate ($\dot{N}_{NS}\sim 1$ per
century) in the Galaxy. The age of the Gould Belt is estimated to
be $\sim$30 Myr old and the pulsars are born at a rate of $\sim
20$ Myr$^{-1}$ (Grenier 2000).

\item{2} The Gould Belt has an ellipsoidal shaped ring with
semi-major and minor axes equal to 500pc and 340pc respectively.
The Sun is displaced from the center of Gould Belt about 200pc
towards $l=130^{\circ}$(Guillout et al. 1998). On the other hand,
there are other possible interpretations of the geometry of the
Gould Belt. For examples, Olano (1982) and Moreno et al. (1999)
have given smaller size for the Gould Belt. However, we will show
that most $\gamma$-ray pulsars originated from the Gould Belt are
mature pulsars with age nearly 1 million years. They already move
very far away from the Gould Belt. The most important point is our
solar system is enclosed in the Gould Belt, the exact dimensions
of the Gould Belt is not very crucial in our problem.

\item{3} The initial position for each pulsar in the Galaxy is
estimated from the distributions $\rho_z(z)=(1/z_{\rm exp})\rm
exp(-|z|/z_{exp})$ and $\rho_R(R)=(a_R/R^2_{\rm exp})$ $R~\rm
exp(-R/R_{exp})$,
 where $z$ is the distance from the Galactic plane, $R$ is the distance
from the Galactic center, $z_{exp}=75$ pc, $a_R=[1-e^{-R_{\rm
max}/R_{\rm exp}}(1+R_{\rm max}/R_{\rm exp})]^{-1}$,
 $R_{\rm exp}=4.5$ kpc and $R_{\rm max}=20$ kpc (Paczynski 1990; Sturner
\& Dermer 1996). But the initial position of each pulsar is
assumed to be born uniformly inside the Gould Belt.

\item{4} The initial magnetic fields are distributed as a Gaussian
in $logB$ with a mean value of 12.5 and a dispersion of 0.3. Since
the majority $\gamma$-ray pulsars are younger than 3 million years
old and the field does not decay in 10 Myr (Bhattacharya et al.
1992). So we ignore any field decay for these rotation-powered
pulsars.

\item{5}  The initial period is chosen to be $P_0$=10ms and the
period at time $t$ is Given by
$P(t)=(P_0+1.95\times10^{-39}B^2t)^{1/2}$. We would like to remark
that the initial period is not very crucial for our problem
because the typical age of $\gamma$-ray pulsars is of order of
million years old.

\item{6} The initial velocity of each pulsar is the vector sum of
the circular rotation velocity at the birth location and random
velocity from the supernova explosion(Paczynski 1990). The
circular velocity is determined by Galactic gravitational
potential and the random velocity is distributed as a Maxwellian
distribution with a dispersion of three dimensional velocity =
$\sqrt{3} \times 100$km/s (Lorimer et al. 1997). Furthermore, the
pulsar position at time $t$ is determined following its motion in
the Galactic gravitational potential. Using the equations given by
Paczynski (1990) for given initial velocity, the orbit
integrations are performed by using the 4th order Runge Kutta
method with variable time step \cite{press92} on the variables
$R$, $V_R$, $z$, $V_Z$ and $\phi$. Then the sky position and the
distance of the simulated pulsar can be calculated.

\item{7} The inclination angle ($\alpha$) of each pulsar is chosen
randomly from a uniform distribution (Biggs 1990).

\item{8}  The following radio selection effects are used. The
pulsar must satisfy that its radio flux is greater than the radio
survey flux threshold and its broadened pulse width is less than
the rotation period (e.g. Sturner \& Dermer 1996). {\bf{We
calculate the 400 MHz radio luminosity, $L_{400}$, of each model
pulsar using the following distribution given by (Narayan \&
Ostriker 1990) $\rho_{L_{400}}(P,
\dot{P})=0.5\lambda^2e^{-\lambda} $, where
$\lambda=3.6(\log(L_{400}/<L_{400}>)+1.8)$, $\log <L_{400}>=6.64+
(1/3)\log(\dot{P}/P^3)$, and $L_{400}$ is in units of mJy kpc$^2$.
The minimum detectable average flux density, $S_{min}$, of a
pulsar's radio survey  is estimated by using the method of Sturner
\& Dermer (1996) (also see Cheng \& Zhang 1998)}}.
 The pulsar
which satisfies $L_{400}/d^2\ge S_{\rm min}$ is considered to be a
radio-detectable pulsar, where $L_{400}$ is the radio luminosity
at 400 MHz and $d$ is the distance to the pulsar. The radio
beaming fraction can be expressed as (Emmering \& Chevalier 1989)
$f_r(\omega)=(1-\cos\omega)+(\pi/2-\omega)\sin\omega$, where
$\omega=6^{\circ}.2\times P^{-1/2}$ (e.g. Biggs, 1990) is the
half-angle of the radio emission cone. Then, following Emmering \&
Chevalier (1989), a sample pulsar with a given period $P$ is
chosen in one out of $f_r(P)^{-1}$  cases using the Monte Carlo
method. {\bf{There are growing evidence that very strong surface
magnetic field exists on the surface of neutron stars. For
examples, absorption/emission line features have been observed in
isolated pulsars/neutron stars,e.g. 1E 1207.4-5209 (Sanwal et al.
2002), PSR 1821-24 (Becker et al. 2003) and RBS1223 (Haberl et al.
2003). These line features imply that the surface magnetic fields
are one to two order of magnitude higher than that of the dipolar
field. In fact, Cheng \& Zhang (1999) have analyzed the X-ray
emission from the polar cap regions of the rotation-powered
pulsars and concluded that there exists a characteristic surface
field with strength $\sim 10^{13}$G. Such strong surface magnetic
field can easily affect the beaming direction of radio wave.}}
Therefore, we assume that the beaming direction of $\gamma$-rays
is independent of the beaming direction of the radio. In general,
it is possible that there is only one beaming, either the radio or
the gamma-ray beam, pointing toward us. In fact this is one of key
differences between outer gap model and polar gap model. The solid
angle of $\gamma$-rays is taken to be 1$sr$.

\item{9} The $\gamma$-ray threshold varies over the sky.
Yadigaroglu  \& Romani (1995) used a flux threshold of $3\times
10^{-10}$ erg cm$^{-2}$s$^{-1}$, which can compare to the faintest
5$\sigma$ sources in the first EGRET catalog(Fichtel et al. 1994).
However, in the third EGRET catalog, the faintest source in the
catalog with significance $\sqrt{TS}\geq$4 has a photon flux of
$(6.2 \pm 1.7)\times 10^{-8}$ cm$^{-2}$s$^{-1}$ (Hartman et al.
1999). Gonthier et al. (2002) have argued that this threshold
could be reduced for $\mid b \mid > 10^{\circ}$. In our analysis,
we include the criterion of the likelihood $\sqrt{TS}\geq$5 ($\sim
5 \sigma$) which corresponds to the energy threshold of
$S^{th}_{\gamma}(E_{\gamma}>100MeV) \geq 1.2\times 10^{-10}$ erg
cm$^{-2}$s$^{-1}$ for $\mid b \mid < 10^{\circ}$ but decreases the
threshold to $S^{th}_{\gamma}(E_{\gamma}>100MeV) \geq 7.0\times
10^{-11}$ erg cm$^{-2}$s$^{-1}$ for $\mid b \mid > 10^{\circ}$
(Gonthier et al. 2002).
\end{description}

\section{Simulation results and Discussion}

{\bf{Following the procedure described in section 4, we perform
Monte Carlo simulation of Galactic pulsars born during past
$3\times 10^7$ yrs. We use the code of Cheng \& Zhang (1998) (also
see Zhang et al. 2000) in our simulation, in which Parkes
Multibeam survey is not included}}. We summarize the various
components of simulated $\gamma$-ray pulsars in table 1. Our
results suggest that there should be more unidentified EGRET
$\gamma$-ray sources identified as radio pulsars. In particular,
the Parkes multi-beam pulsar survey is expected to discovery more
radio pulsars (Manchester et al. 2001). Torres et al. (2001) have
found five new possible radio pulsar-Unidentified source
associations. Most recently, Kramer et al. (2003) have used the
newly release survey results to correlate the unidentified EGRET
$\gamma$-ray sources and, have found more than 35 coincidences and
around 20 probable associations. It is important to note that the
number in each box of table 1 is sensitive to the input
parameters, e.g. detection threshold, birth rate, initial
distributions of pulsars etc., for a given theoretical model.
However, their ratios are less sensitive to most input parameters
except their relative birth rate. It is interested to note that
the $\gamma$-ray pulsars in the Gould Belt contribute significant
fraction of total $\gamma$-ray pulsars for $\mid b \mid >
5^{\circ}$ and become unimportant for $\mid b \mid < 5^{\circ}$.
This is simply because the solar system is enclosed by the Gould
Belt.

In Fig. 5a-e, we compare our simulated distributions of period,
period derivative, magnetic field, distance and energy flux with
the observed data of 8 known radio-loud $\gamma$-ray pulsars (i.e.
Crab, Vela, Geminga, PSR 0656+14, PSR 1046-58, PSR 1055-52, PSR
1509-58, PSR 1706-44, PSR 1951+32 ) using KS test. Although PSR
0656+14 and PSR 1509-58 are confirmed in $\gamma$-ray band, there
are insufficient $\gamma$-ray photons to provide the information
of pulsed fraction. So we did not include these pulsars in the
cumulative plot of energy flux. The maximum deviations of period,
period derivative, magnetic field, distance and $\gamma$-ray
energy flux distributions from the observed distributions are
0.36, 0.32, 0.25, 0.33 and 0.21 respectively. It can be seen that
four of five accumulative distributions cannot be rejected at
better 80\% confidence level, and period cumulative distribution
cannot be rejected at better than 90\% confidence level. Therefore
we conclude that the model results do not conflict with the
observed data of $\gamma$-ray pulsars.

In Fig. 6a-f, we plot the normalized distributions of period,
magnetic field, age, distance, inclination angle and the
fractional size of outer gap. The solid lines and the dashed lines
are $\gamma$-ray pulsars located at $\mid b \mid < 5^{\circ}$ and
$\mid b \mid> 5^{\circ}$ respectively. We can clearly see that
there are two classes of $\gamma$-ray pulsars. $\gamma$-ray
pulsars located at $\mid b \mid < 5^{\circ}$ have shorter periods,
younger, larger inclination angles and smaller outer gap size in
comparing with $\gamma$-ray pulsars located at $\mid b \mid >
5^{\circ}$. As we have mentioned that $\gamma$-ray pulsars in high
galactic latitude are dominated by mature pulsars, which are old
enough to move up the high latitude and to evolve to longer
periods. Young $\gamma$-ray pulsars will be dominated in the
galactic plane, which have stronger $\gamma$-ray luminosity,
larger inclination angle, shorter period and larger magnetic
field. In Fig. 6, $\gamma$-ray pulsars in high galactic latitude
are actually closer than those in galactic plane. It is because
they are older and weaker $\gamma$-ray pulsars, so they must be
near otherwise they cannot be detected.

In Fig. 7, we plot $L_{\gamma}$ versus $L_{sd}$, where the upper
panel is $\gamma$-ray pulsars located at $\mid b \mid < 5^{\circ}$
and lower panel is $\gamma$-ray pulsars located at $\mid b \mid >
5^{\circ}$. We can see that there are two distinctive regions. For
$L_{sd} > 3 \times 10^{34}$ erg/s, the relation between
$L_{\gamma}$ and $L_{sd}$ is rather scattered. However, for
$L_{sd} < 3 \times 10^{34}$ erg/s, $L_{\gamma}$ is proportional to
$L_{sd}$. In particular, the $\gamma$-ray pulsars in high
latitude, most pulsars satisfy $L_{\gamma} \propto L_{sd}$, which
means $f \sim 1$. This is supported by Fig. 6f, in which the
fractional sizes of outer gap $f$ for $\gamma$-ray pulsars located
at $\mid b \mid > 5^{\circ}$ are all close to unity whereas the
fractional sizes of outer gap for $\gamma$-ray pulsars located at
$\mid b \mid < 5^{\circ}$ have a wide distribution. If $L_{\gamma}
\propto L^{\beta}_{sd}$ is used to fit all $\gamma$-ray pulsars,
then $\beta \sim 0.3$ for $\gamma$-ray pulsars located at $\mid b
\mid < 5^{\circ}$ and $\beta \sim 0.6$ for $\gamma$-ray pulsars
located at $\mid b \mid
> 5^{\circ}$ respectively.  Since the outer
gap size is a function of $B, P and \alpha$. For young pulsars,
there are more combination of these three parameters to make the
outer gap size less than unity. On the other hand, for old
$\gamma$-ray pulsars their periods are already so long that there
are not much room for other two parameters to vary the outer gap
size away from unity. The spatial distributions of $\gamma$-ray
pulsars are given in Fig. 8.

In Fig. 9a-d, we compare distributions of $\gamma$-ray pulsars
from the Galaxy (dotted lines) and from the Gould Belt (solid
lines). We can see that the properties of $\gamma$-ray pulsars
from two sources are very similar except $\gamma$-ray pulsars from
the Gould Belt have lower population and closer. These two
properties results from the lower birth rate in the Gould Belt and
the solar system is inside the Gould Belt.

\section{\bf{Conclusion and discussion}}

We have studied the $\gamma$-ray emission properties of pulsars in
the Galaxy as well as in the Gould Belt by using a new
self-consistent outer gap model, which includes the effects of
inclination angle and average properties of outer gap. We have
found that this new model can produce more mature $\gamma$-ray
pulsars, which have longer period, smaller inclination angle and
typical age of one million years old, than the old model. In fact
the mature $\gamma$-ray pulsars dominate in the high galactic
latitude. The spectra of these mature pulsars are significantly
softer and weaker than the young pulsars in the galactic plane. It
is because the spectral hardness is roughly determined by the
typical photon energy given in equation (1), which characterizes
the position of spectral break. The explicit form of how the
typical photon energy depends on the inclination angle is given in
equations (28) and (38) of Zhang et al. (2004). Although the form
is quite complicated, we can roughly understand why smaller
inclination angle gives smaller typical photon energy and hence
{\bf {softer}} spectrum by equation (1). It is because the typical
position of the outer gap is the null surface where becomes larger
for smaller inclination angle. From the simulation results, it is
not clear why $\gamma$-ray pulsars in high latitude should have
smaller inclination angle. It is because all of these high
latitude pulsars are older pulsars, which have longer period.
Without the compensation of the effect of inclination angle, the
fractional size of the outer gap $f$ is larger than unity ( cf.
equation 3). In order to maintain the outer gap size less than
unity, $\eta$ in equation (4), which contains the effect of
inclination angle must be small. In fact $\eta$ becomes small for
small inclination angle (cf. Fig. 1 of Zhang et al. 2004).
Therefore high latitude gamma-ray pulsars tend to have smaller
inclination angle and harder spectrum. On the other, this seems to
imply that the confirmed radio-loud $\gamma$-ray pulsars in the
galactic plane should have larger inclination. This statistical
correlation has not yet been established. But we would like to
point out that the values of inclination angle of pulsars are very
difficult to be measured accurately and there are only 7 confirmed
$\gamma$-ray pulsars. So it is very difficult to find such
correlation in radio-loud $\gamma$-ray pulsars. However, we do
believe that this correlation should be there when more radio-loud
$\gamma$-ray pulsars are discovered in the galactic plane.

We have also used a Monte Carlo method to simulate the properties
of the $\gamma$-ray pulsar population in the Galaxy as well as in
the nearby Gould Belt in terms of the revised outer gap models.
The initial magnetic field, spatial and velocity distributions of
the neutron star at birth which are obtained by the radio pulsar
statistical studies have been used in our simulations. We have
obtained the spatial, distance, period, age, magnetic field,
inclination angle, photon flux distributions of the radio-loud and
radio-quiet $\gamma$-ray pulsars. We find that the properties of
$\gamma$-ray pulsars between $\mid b \mid < 5^{\circ}$ and $\mid b
\mid > 5^{\circ}$ are very much different. Galactic plane
$\gamma$-ray pulsars are younger, shorter in period and have
larger inclination angles. They satisfy different relation between
$L_{\gamma}$ and $L_{sd}$. The high latitude $\gamma$-ray pulsars
satisfy $L_{\gamma}\propto L^{0.6}_{sd}$ but the galactic plane
pulsars satisfy $L_{\gamma}\propto L_{sd}^{0.3}$.

The present model predicts very similar numbers of $\gamma$-ray
pulsars as old models  but many more high latitude $\gamma$-ray
pulsars than old models (Cheng \& Zhang 1998; Zhang, Zhang \&
Cheng 2000). It is because the new model allows the outer gap can
exist about a few million years old for appropriate combination of
B,P and $\alpha$, so old $\gamma$-ray pulsars can move up to high
galactic latitude. Furthermore, the nearby Gould Belt also
contribute significant number of $\gamma$-ray pulsars in high
latitude. Torres et al. (2003) have predicted that AGILE can
detect more $\gamma$-ray pulsars (for a general review of AGILE
cf. Tavani et al. 2001). Perhaps AGILE can provide some clues to
differentiate polar gap model predictions (Gonthier et al. 2002)
and outer gap model predictions, and GLAST makes the final
verdict. Finally we would like to remark that the better
predictions on how many $\gamma$-ray pulsars should be detected by
AGILE and GLAST should include a better expression of $\gamma$-ray
solid angle, which depends on the inclination angle as well as the
outer gap size. However, in order to have a reliable expression of
$\gamma$-ray solid angle, we need to carry out three-dimensional
model calculation and come up with an approximate expression from
Monte Carlo simulation.

\acknowledgments{We thank an anonymous referee for the very useful
comments. This work is partially supported by a RGC grant of Hong
Kong Government, 'Hundred Talents Program of CAS' and the National
973 Projection of China (NKBRSFG 19990754).}

\clearpage
\begin{table}
\caption{Various Components of Simulated $\gamma$-ray Pulsars }
\begin{tabular}{|p{1.4in}|p{0.6in}|p{0.6in}|p{0.6in}|p{0.6in}|} \hline

    &\multicolumn{2}{c|}{Gould Belt} &\multicolumn{2}{c|}{Galaxy}     \\\hline
    Birth rate  &\multicolumn{2}{c|}{1/50,000yr.} &\multicolumn{2}{c|}{1/100yr.}     \\\hline
     Galactic latitude $|b|$ & $<5^\circ$ & $>5^\circ$ & $<5^\circ$ & $ >5^\circ$\\\hline
    Radio-loud  & 1 & 4  & 12 & 4     \\\hline
    Radio-quiet & 2 & 13 & 40 & 22    \\ \hline

\end{tabular}
\end{table}

\clearpage


\begin{figure}[ht]
\vbox to2.25in{\rule{0pt}{2.25in}} \includegraphics{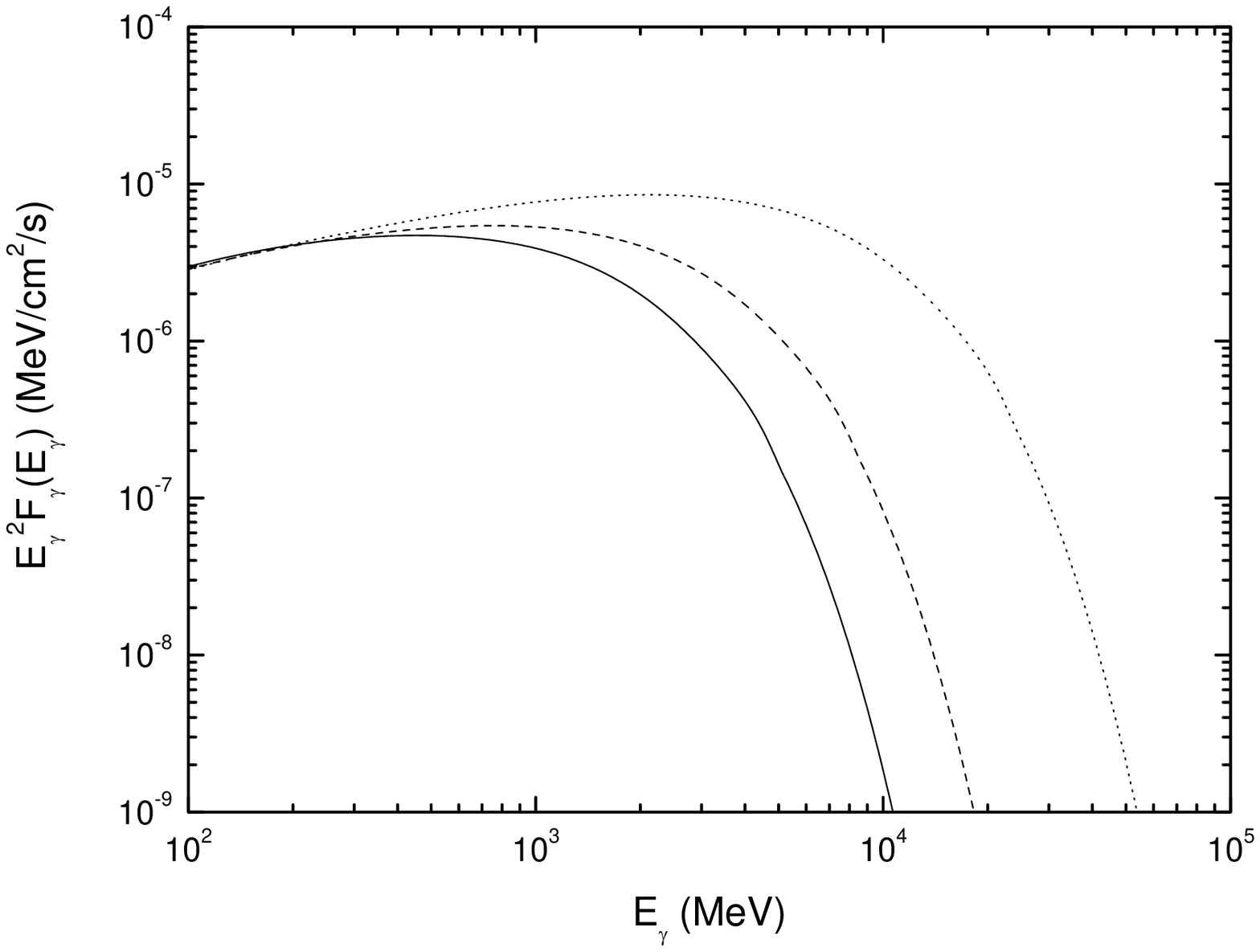} \vspace{0.1in}
\noindent{Fig.~1 Distribution of the energy flux vs energy of the
photon. In this figure, $B_{12}$ has been taken to be 3, $P$ to be
0.4 s and $\alpha $ to be $20^\circ$ (solid line), $40^\circ$
(dashed line) and $60^\circ$ (dotted line). \label{fig1}}
\end{figure}

\begin{figure}[ht]
\vbox to2.5in{\rule{0pt}{2.5in}} \includegraphics{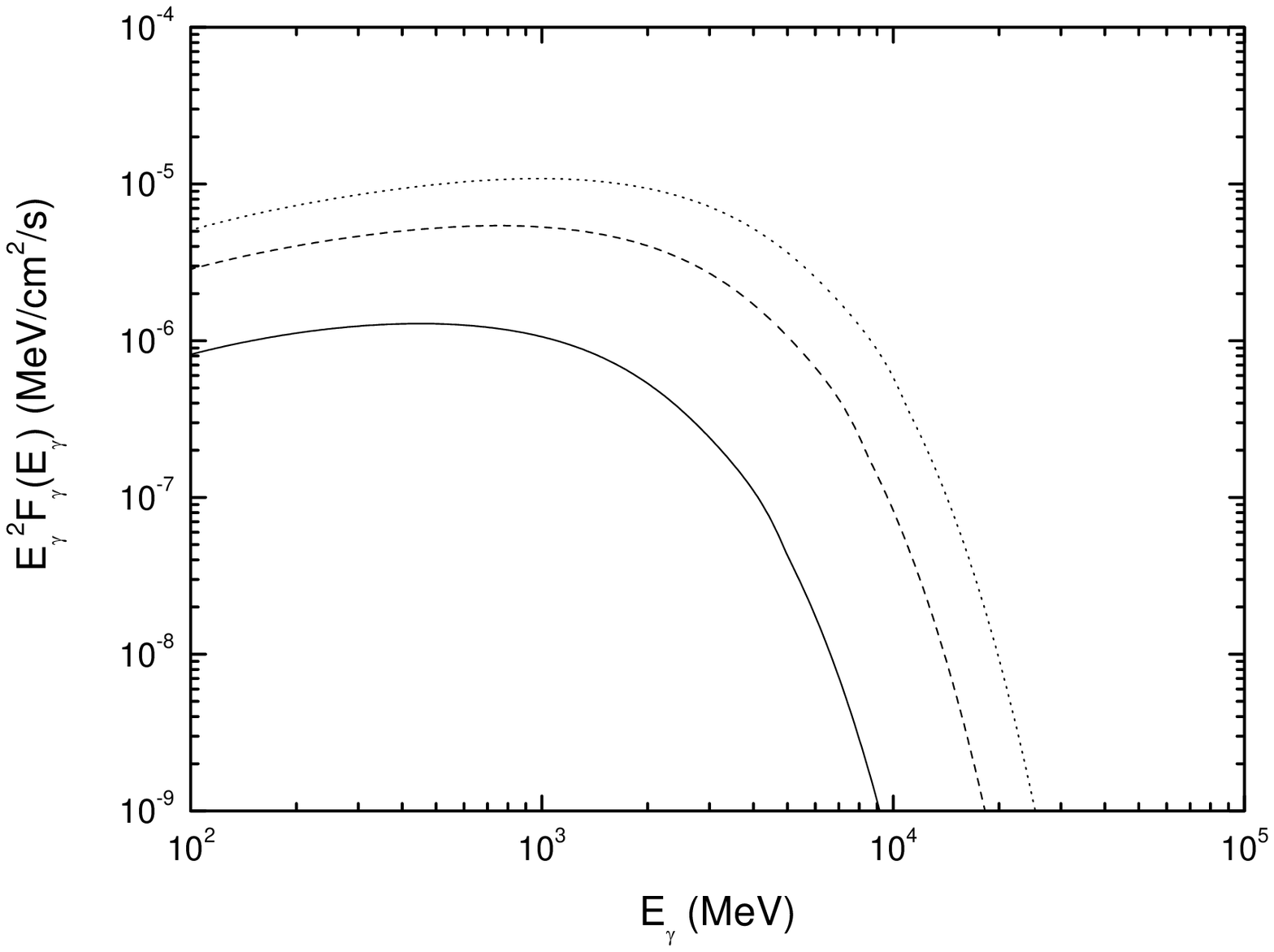} \vspace{0.1in}
\noindent{Fig.~2 Distribution of the energy flux vs energy of the
photon. In this figure, $P$ has taken to be 0.4 s, $\alpha $ to be
$40^\circ$ and $B_{12}$ to be 1 (solid line), 3
 (dashed line) and 5 (dotted line).\label{fig2}}
\end{figure}

\begin{figure}[ht]
\vbox to2.5in{\rule{0pt}{2.5in}} \includegraphics{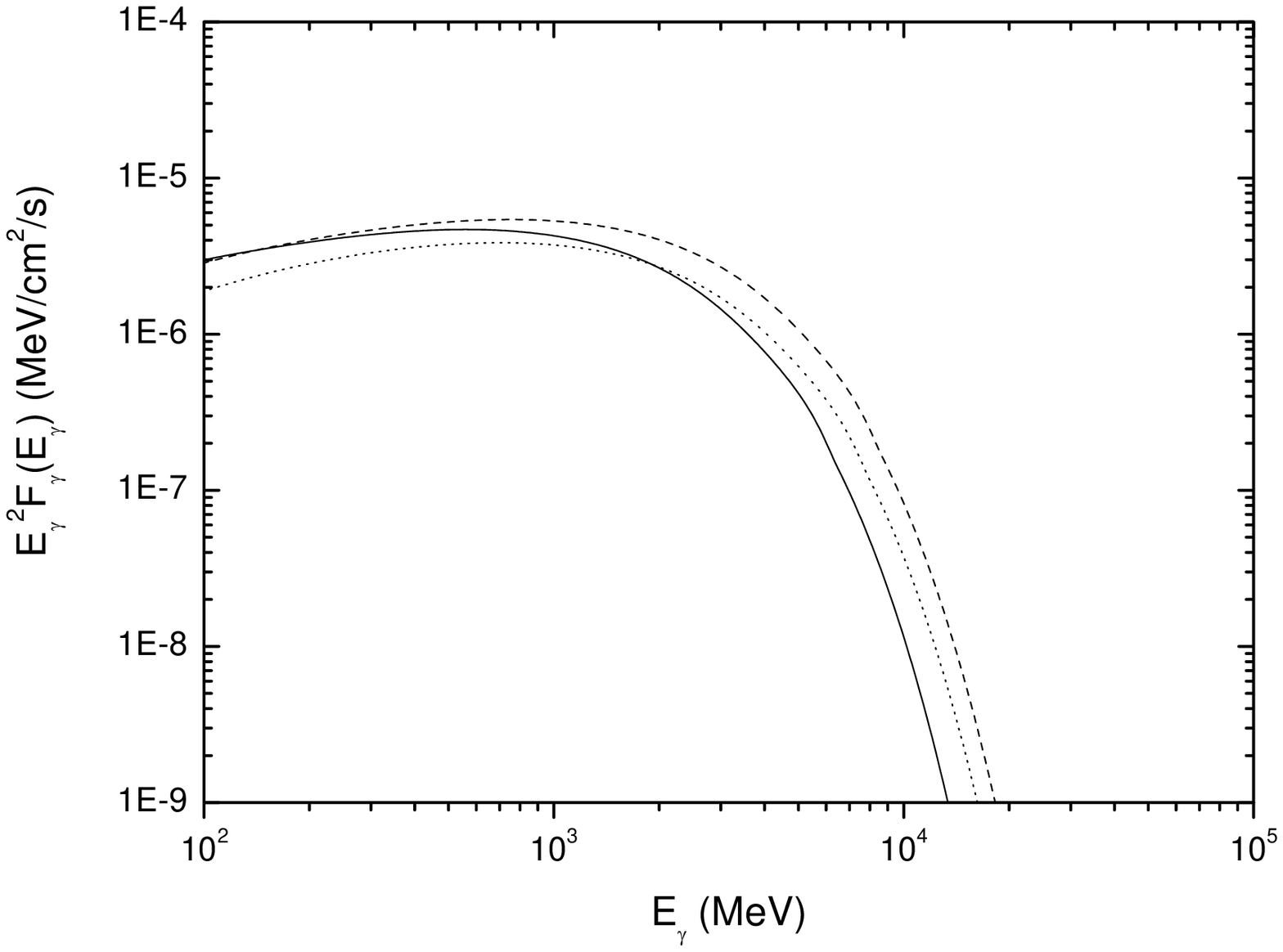} \vspace{0.1in}
\noindent{Fig.~3 Distribution of the energy flux vs energy of the
photon. In this figure, $B_{12}$ has taken to be 3, $\alpha $ to
be $40^\circ$ and $P$ to be 0.2 s (solid line), 0.4 s (dashed
line) and 0.6 s (dotted line).}
\end{figure}

\begin{figure}[ht]
\vbox to2.5in{\rule{0pt}{2.5in}} \includegraphics{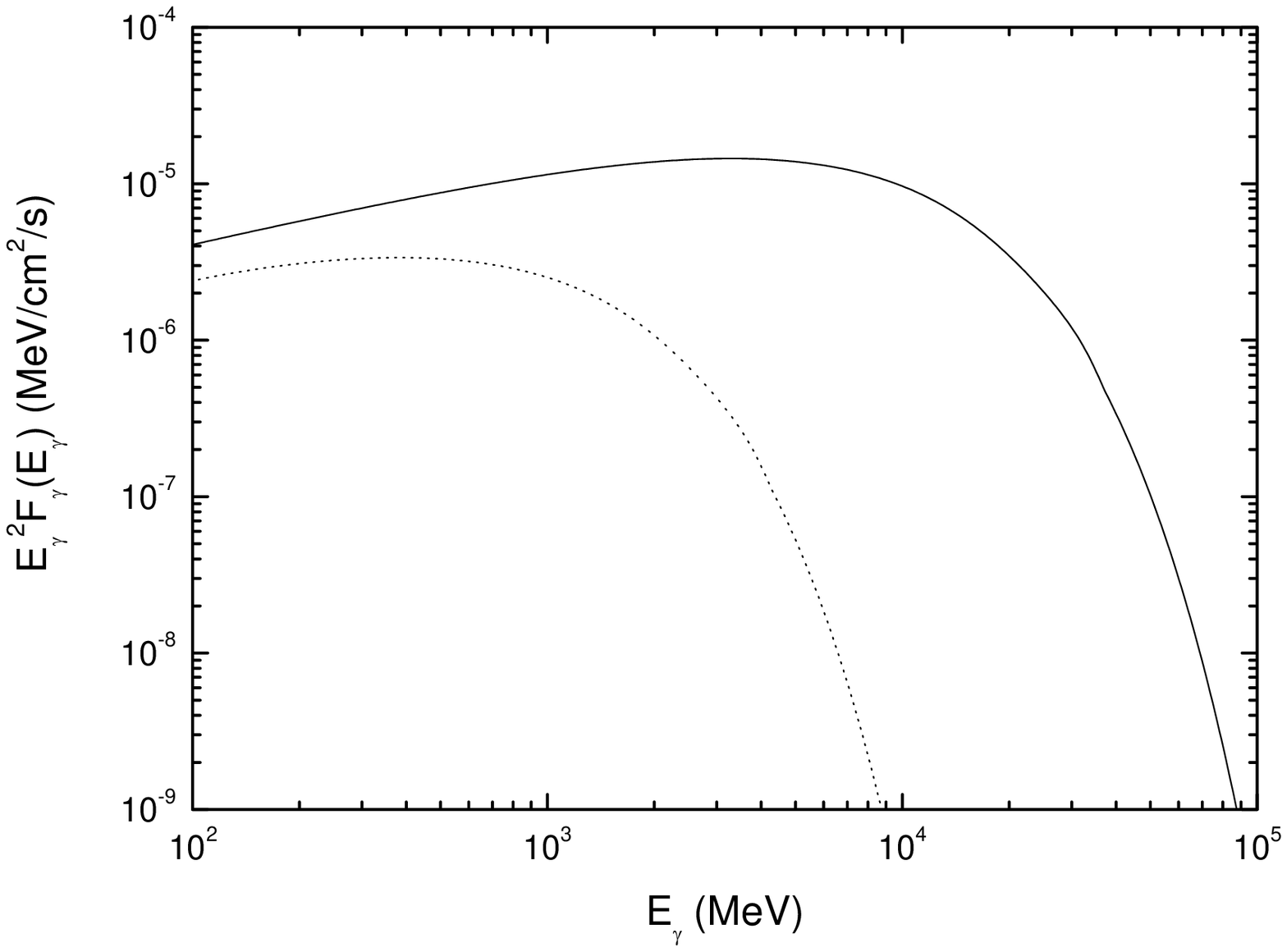} \vspace{0.1in}
\noindent{Fig.~4 Distribution of the energy flux vs energy of the
photon. The solid line is the typical photon spectrum of a
galactic $\gamma$-ray pulsar with $B_{12}=5$, $P=0.1s$ and
$\alpha=70^\circ$, and the dashed line is the typical photon
spectrum of a $\gamma$-ray pulsar in high galactic latitude with
$B_{12}=2$, $P=0.3s$ and $\alpha=20^\circ$ respectively. }
\end{figure}

\begin{figure}[ht]
\vbox to6.2in{\rule{0pt}{6.2in}} \includegraphics{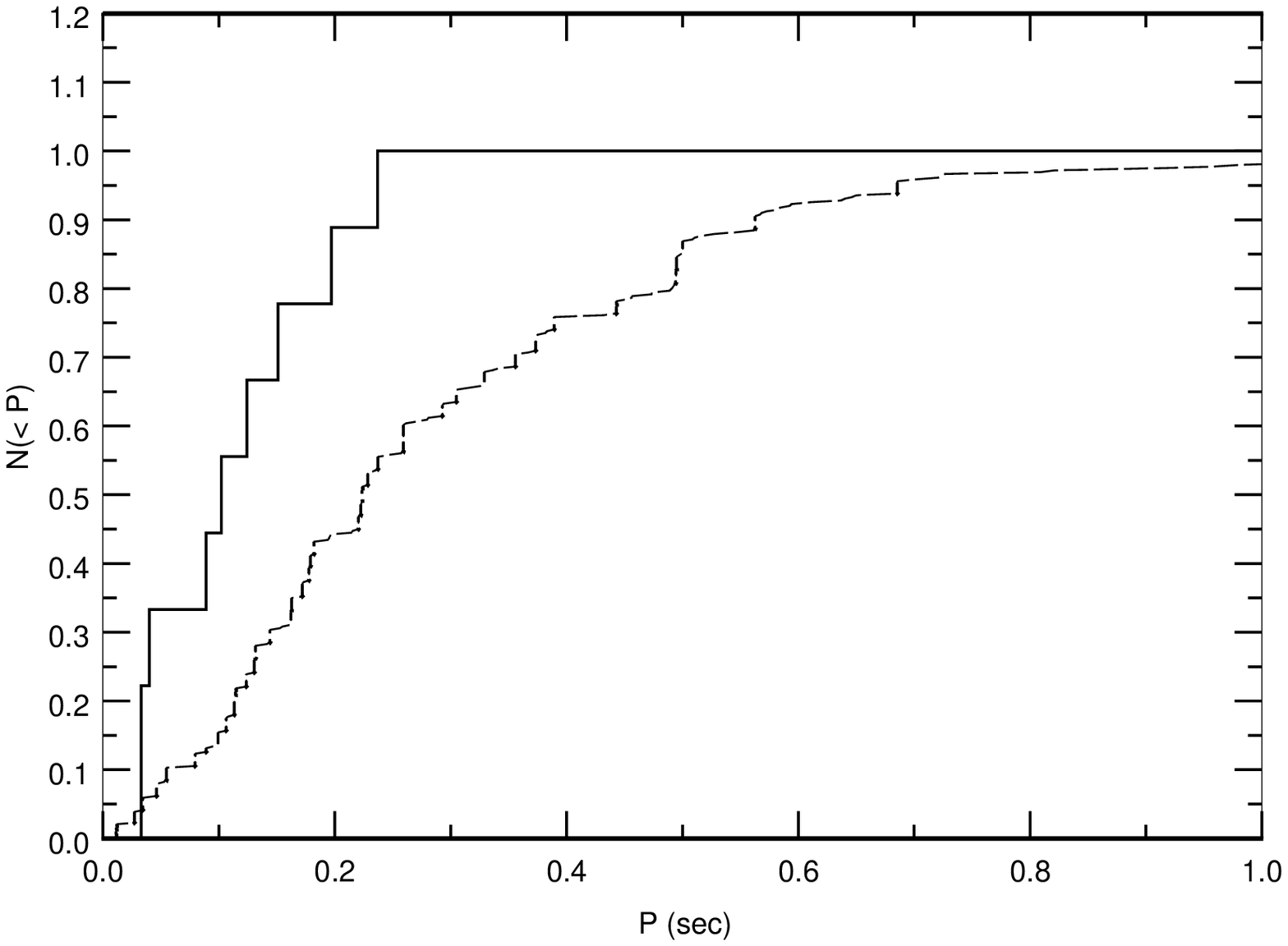}
\includegraphics{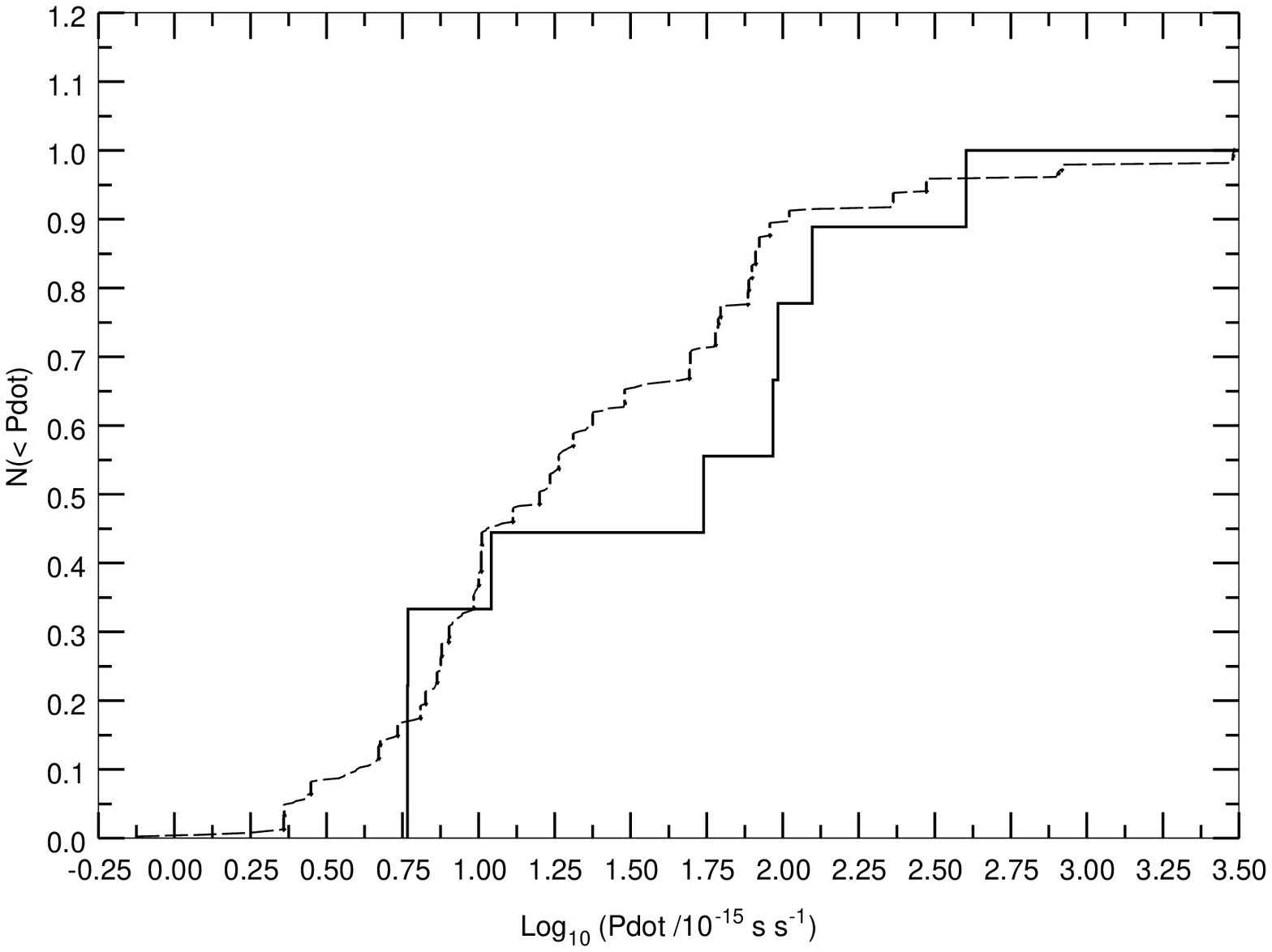} \includegraphics{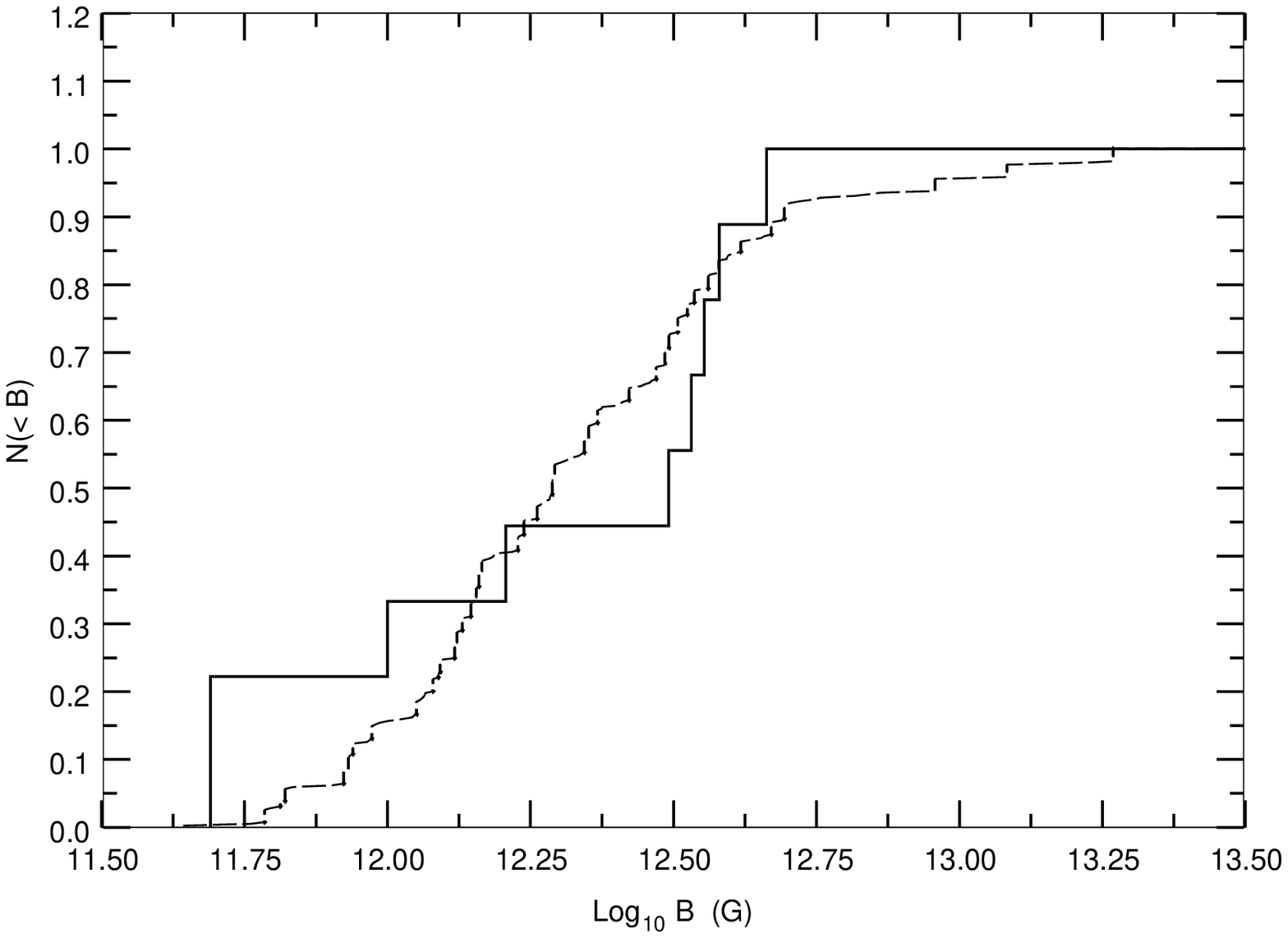} \includegraphics{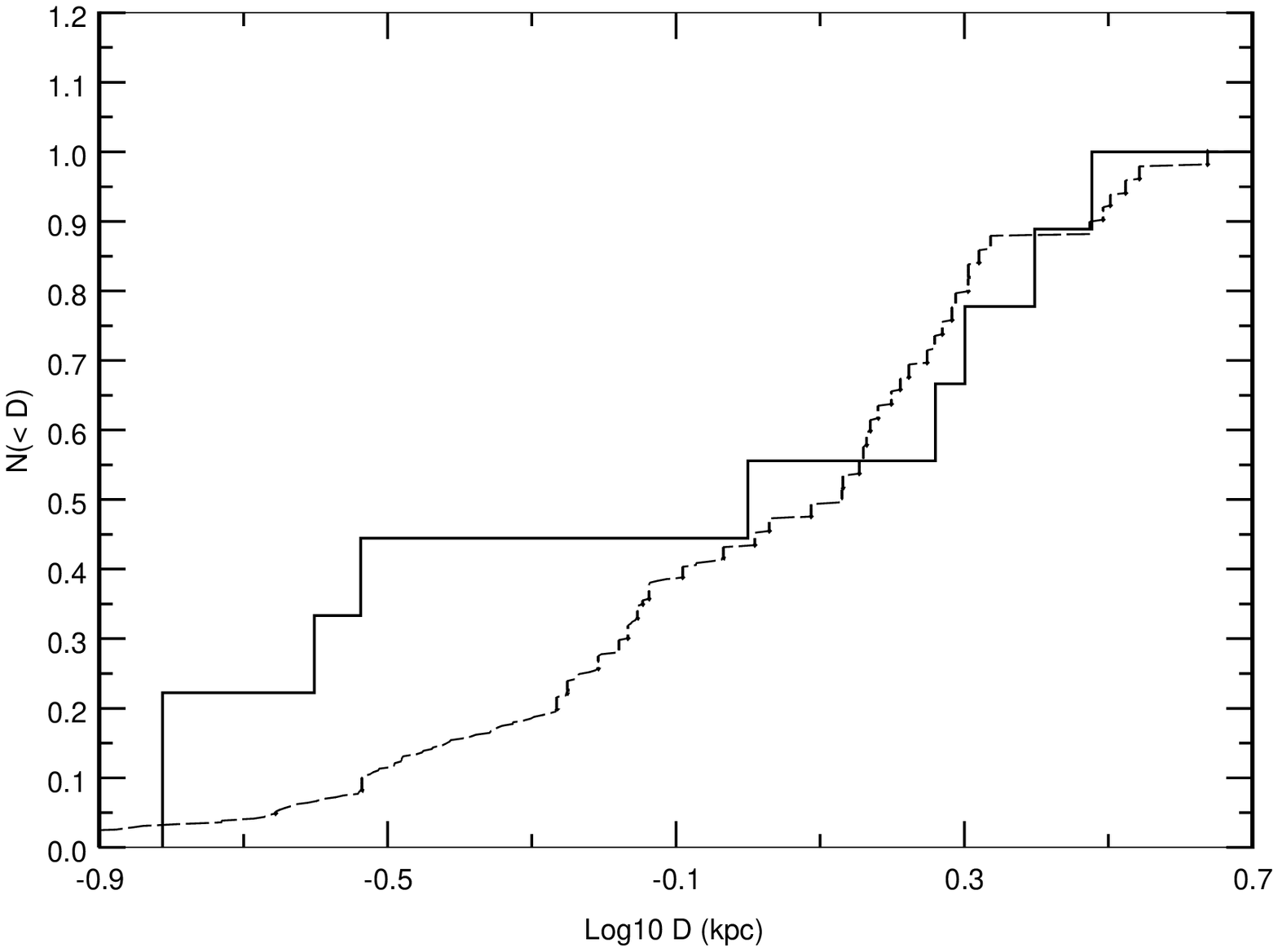}
\includegraphics{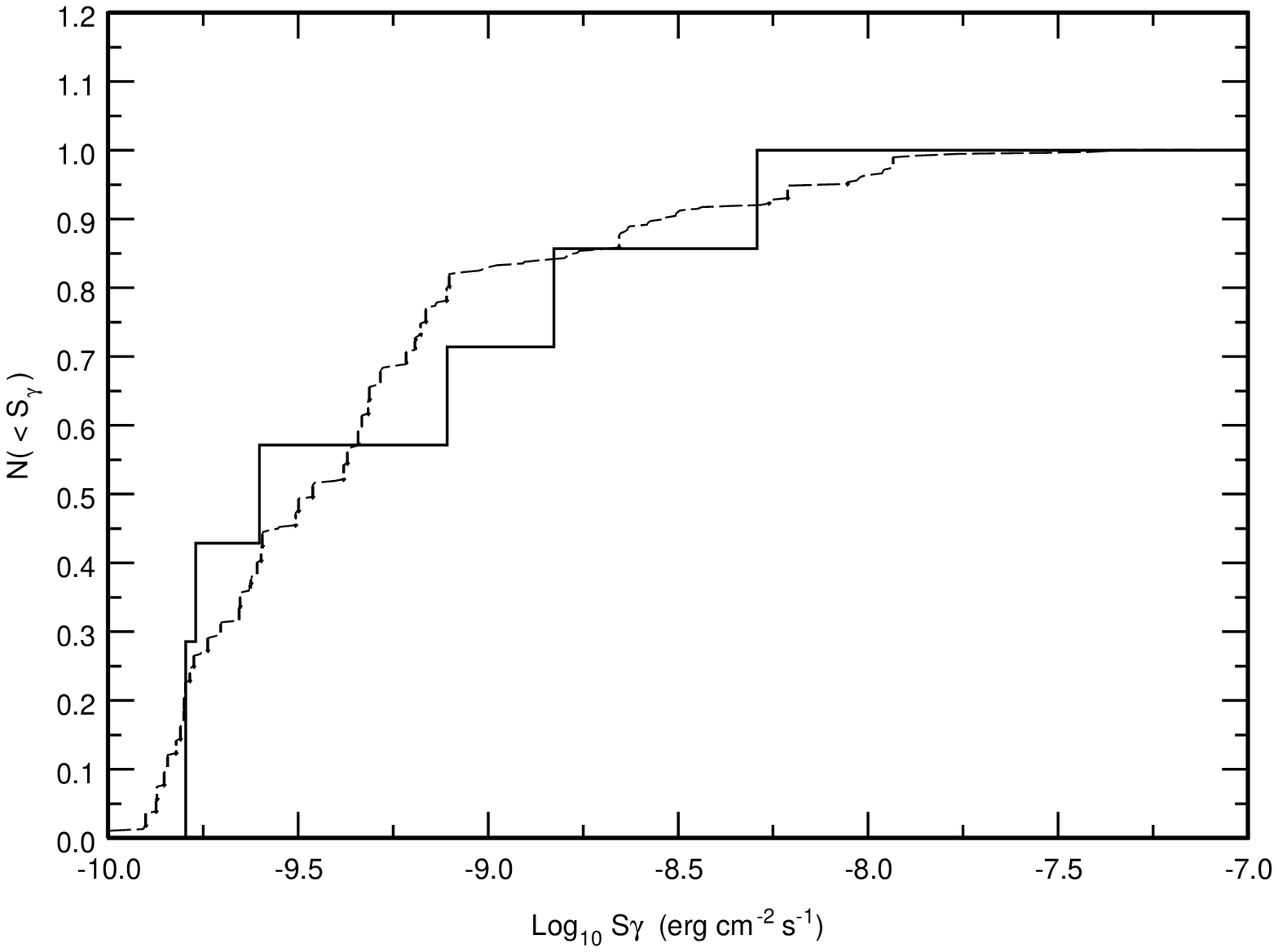}
 \noindent{Fig.~5 Normalized cumulative distributions ({\bf{dashed curves}}) of period, period derivative,
 magnetic field, distance and $\gamma$-ray energy flux of $\gamma$-ray pulsar
 population for our model with radio selection effects. For
 comparison, corresponding distributions ({\bf{solid lines}}) of the observed
 $\gamma$-ray pulsars are also shown. }
\end{figure}

\begin{figure}[ht]
\vbox to6.2in{\rule{0pt}{6.2in}} \includegraphics{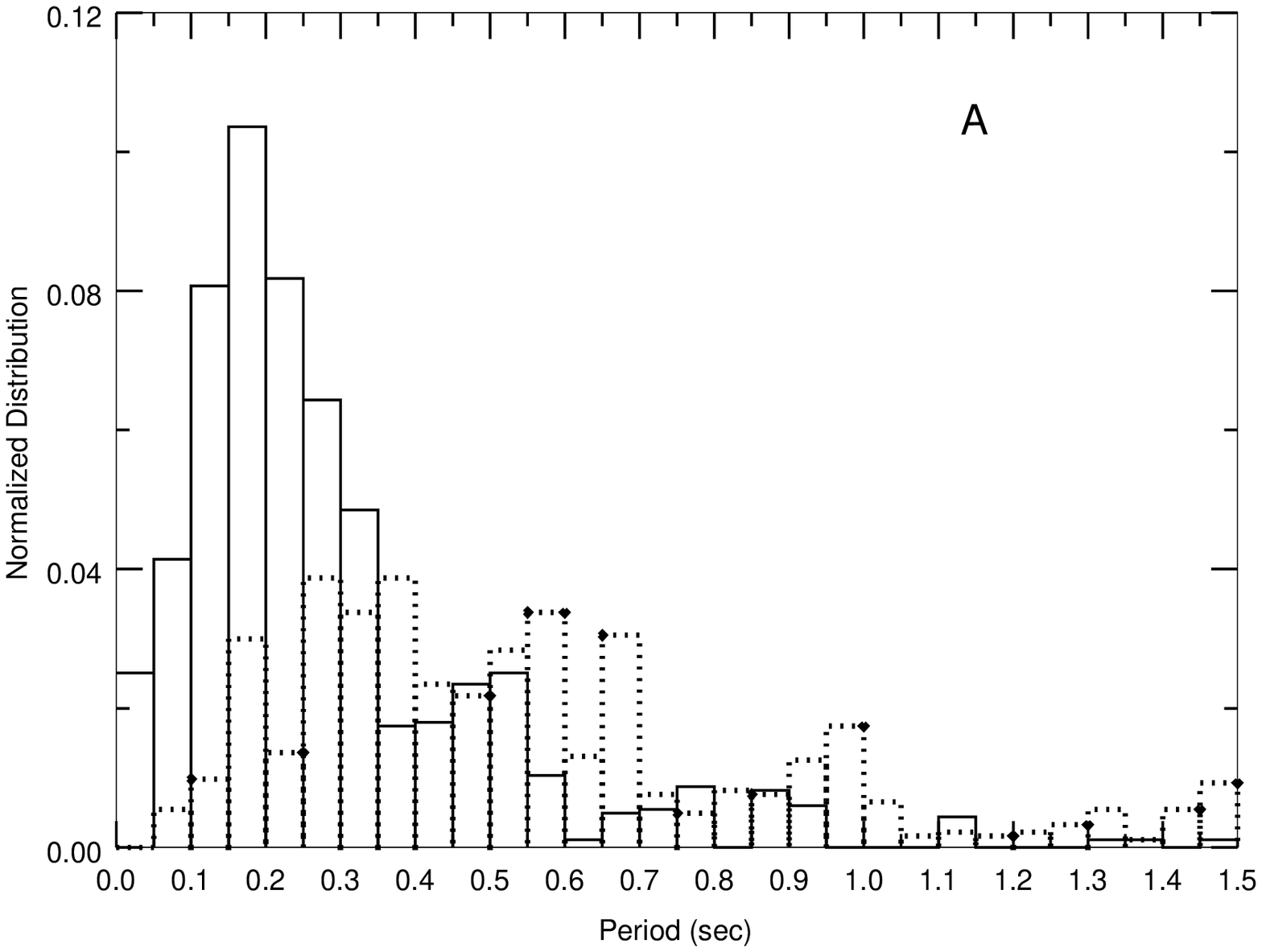}
\includegraphics{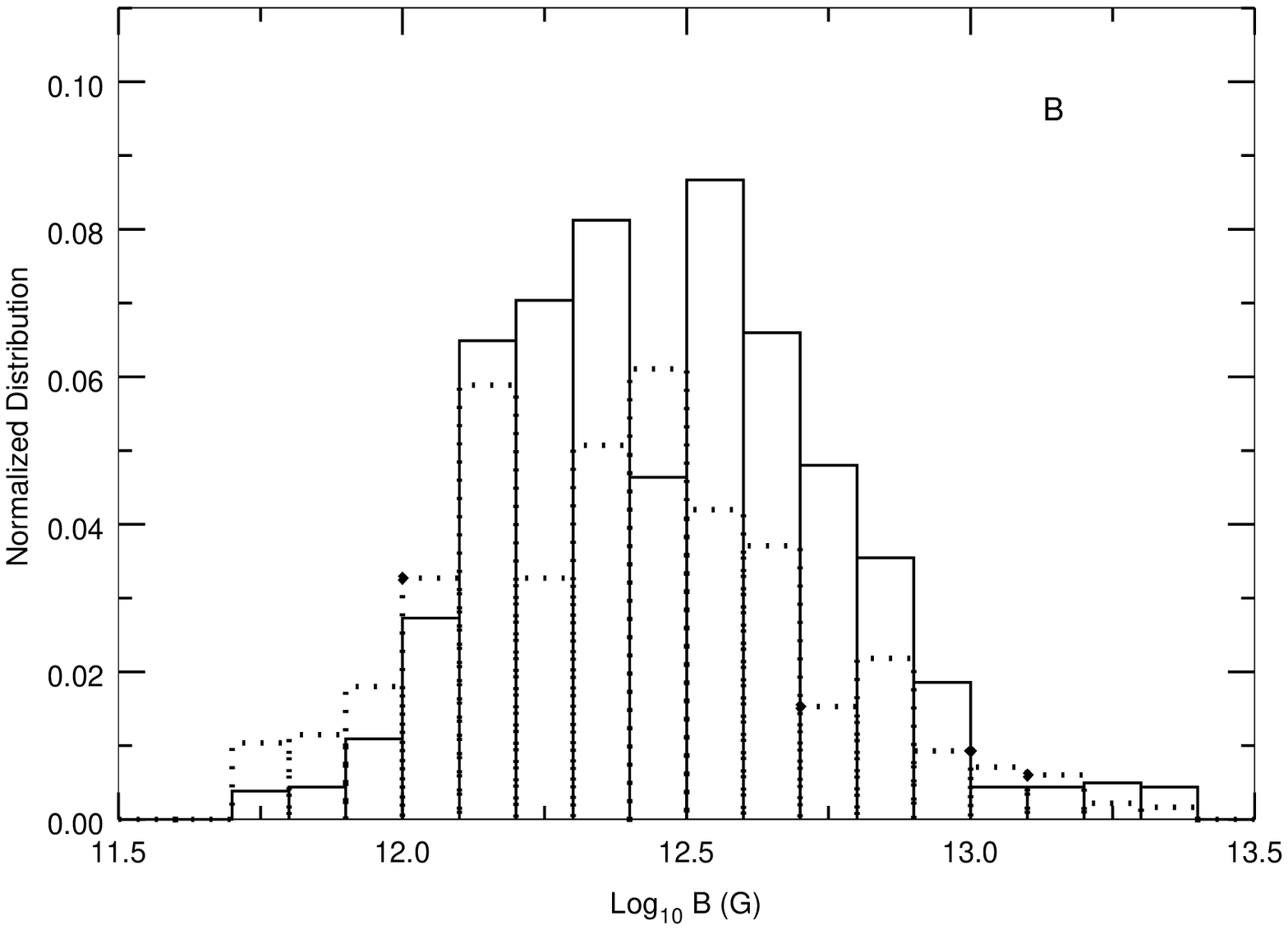} \includegraphics{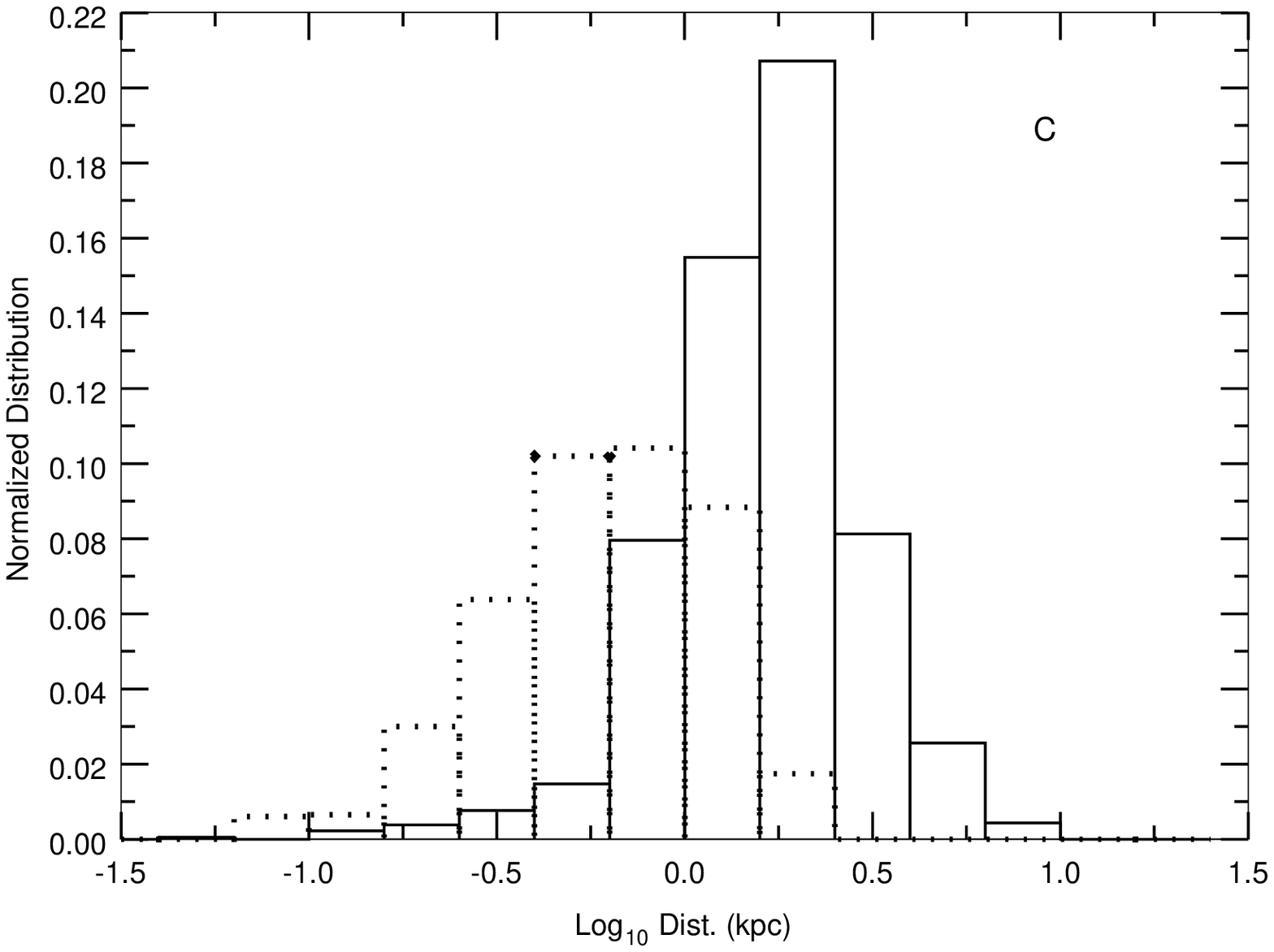} \includegraphics{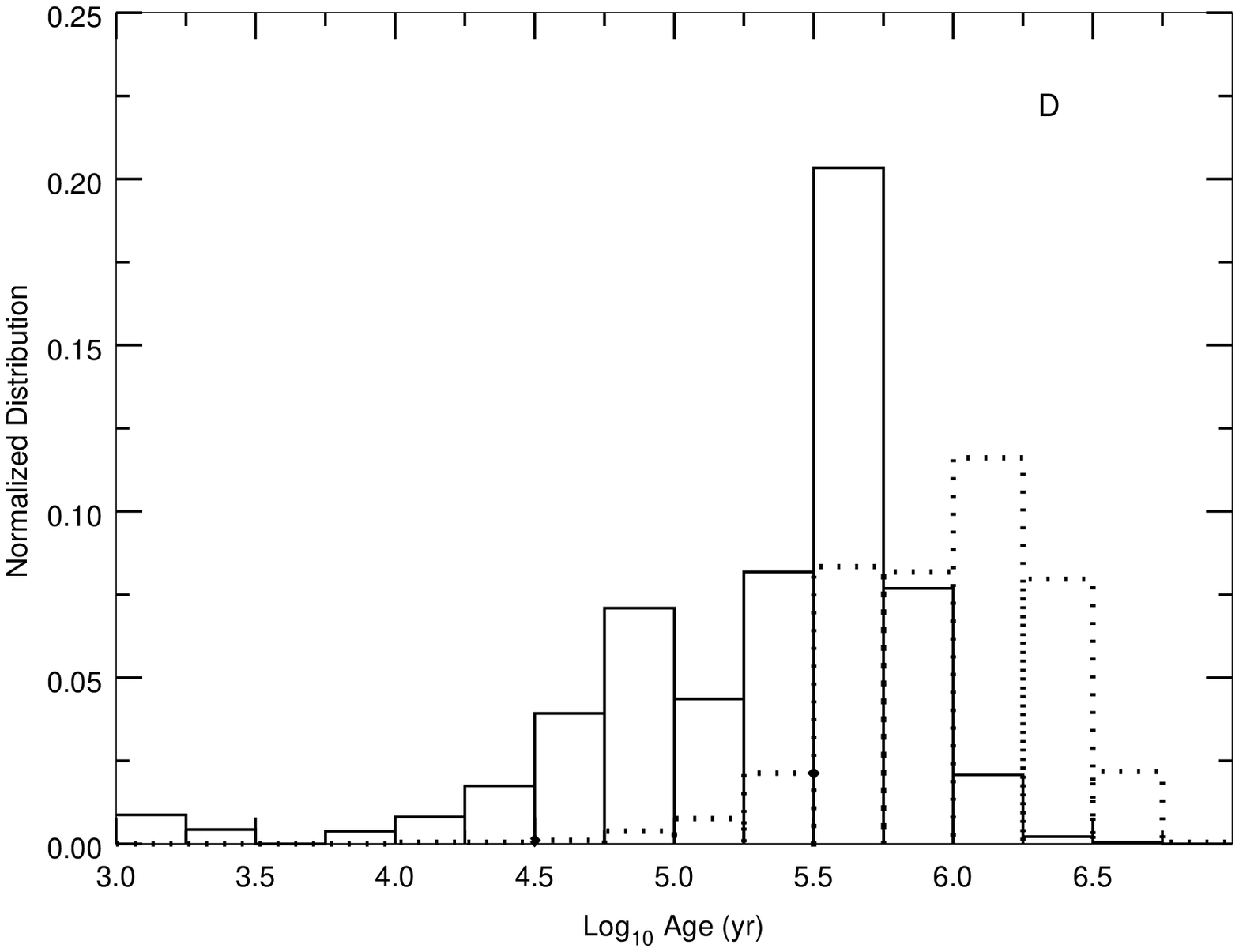}
\includegraphics{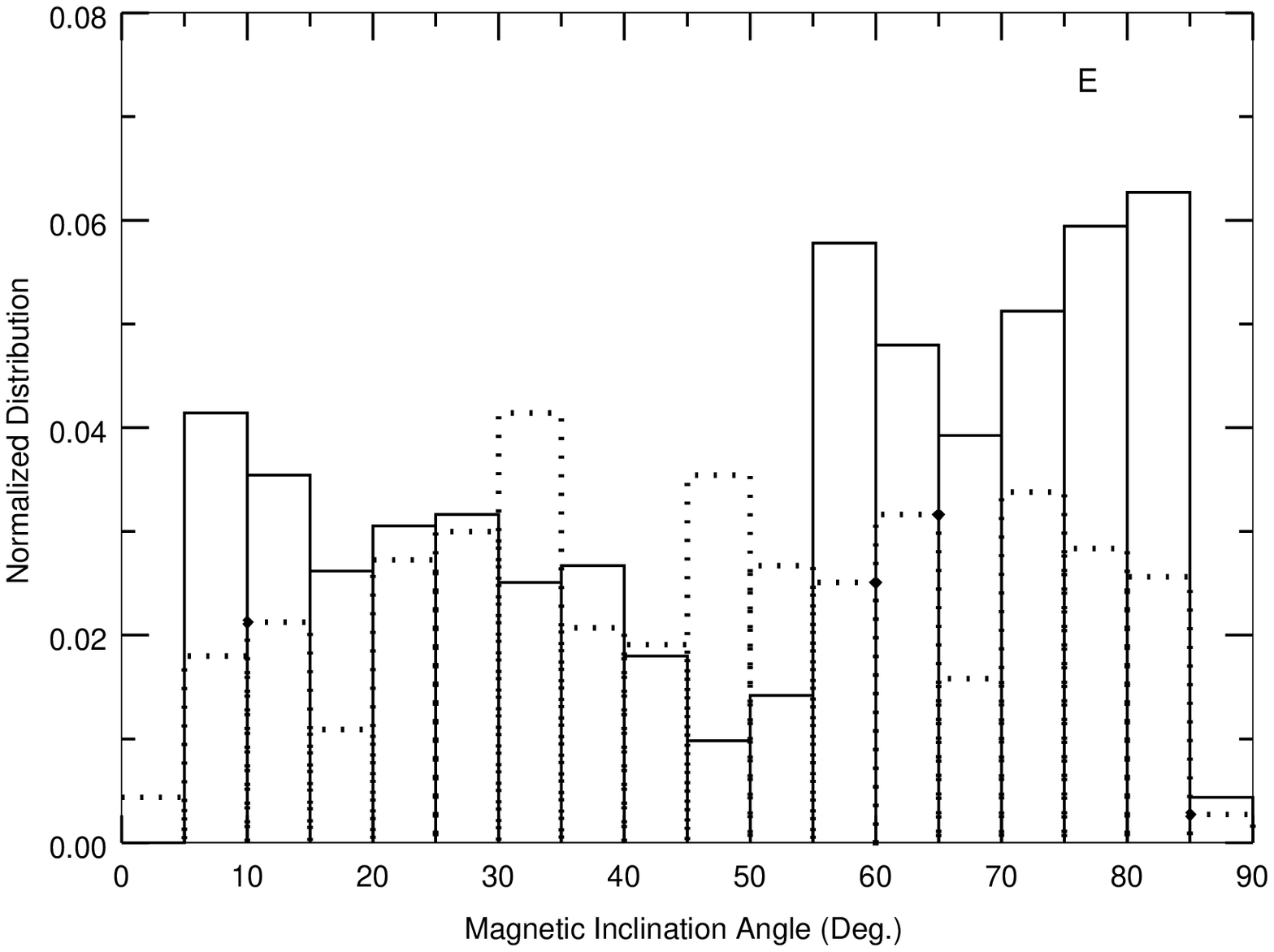}\includegraphics{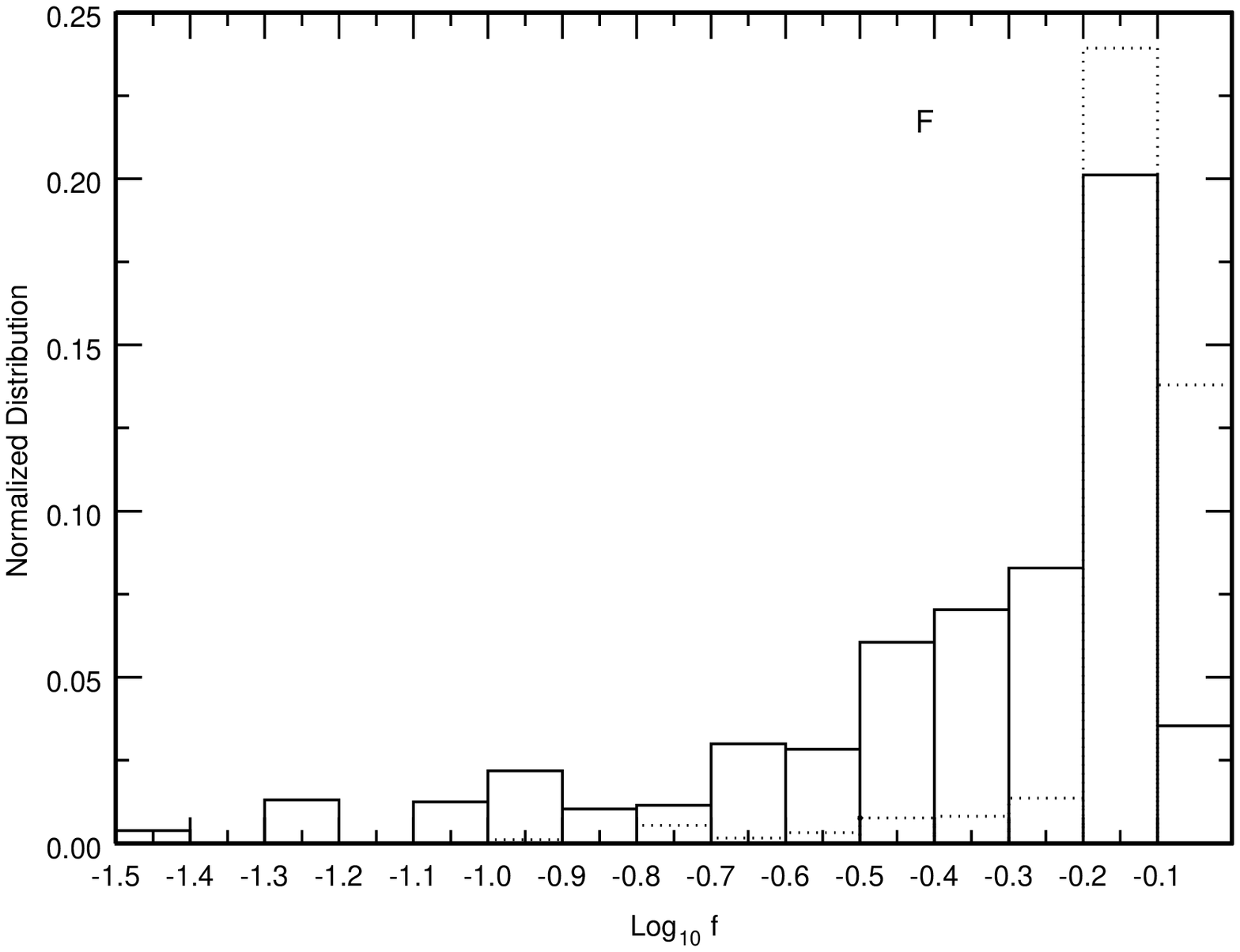}
 \noindent{Fig.~6 Normalized cumulative distributions of period,
 magnetic field, distance, age, inclination angle and the averaged fractional size of outer gap
 of both simulated $\gamma$-ray pulsars in $|b|\le 5^{\circ}$ (solid lines )and in $|b|>5^{\circ}$
 (dashed lines), {\bf{which are labelled  by A, B, C, D, E, F respectively}}. }
\end{figure}

\begin{figure}[ht]

\vbox to5in{\rule{0pt}{5in}} \includegraphics{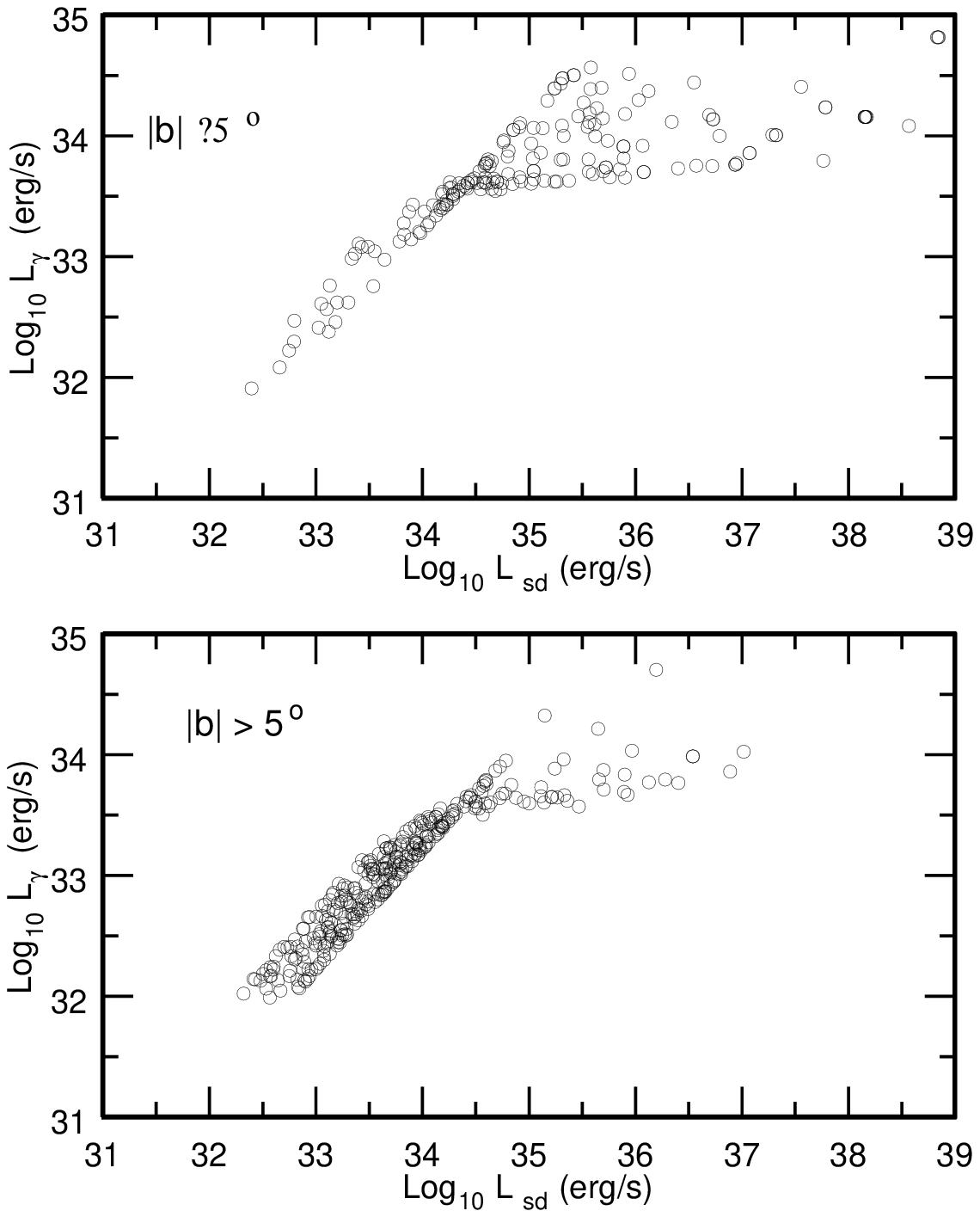} \noindent{Fig.~7 Plot of
$\gamma$-ray luminosity vs. spin-down luminosity for the simulated
$\gamma$-ray pulsars. Upper panel for the case in $|b|\le
5^{\circ}$, and bottom panel for the case in $|b|>5^{\circ}$. }
\end{figure}

\begin{figure}[ht]
\vbox to5in{\rule{0pt}{5in}} \includegraphics{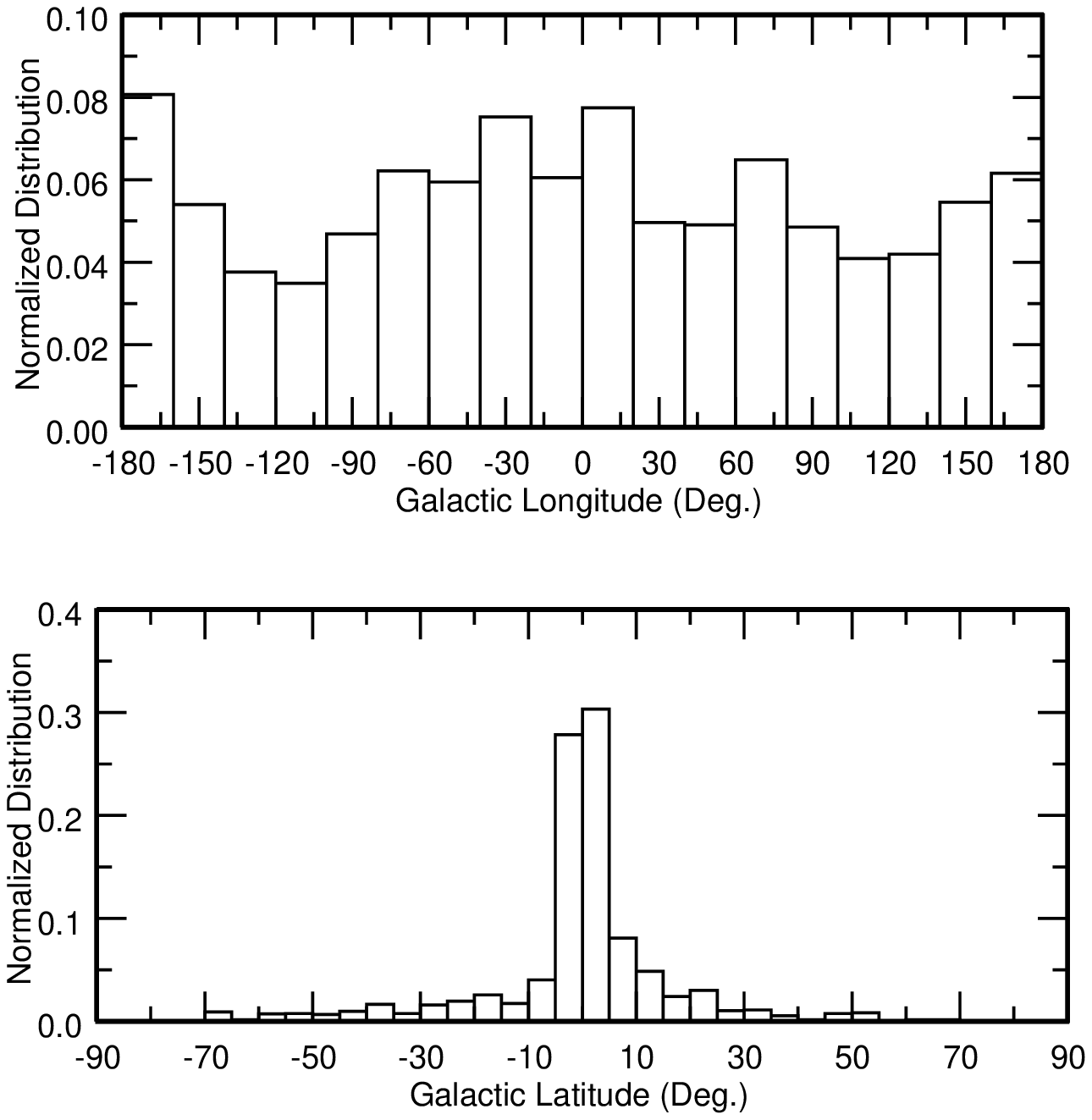} \noindent{Fig.~8 Normalized
distributions of the simulated $\gamma$-ray pulsars in Galactic
longitude and latitude.}
\end{figure}

\begin{figure}[ht]
\vbox to6.2in{\rule{0pt}{6.2in}} \includegraphics{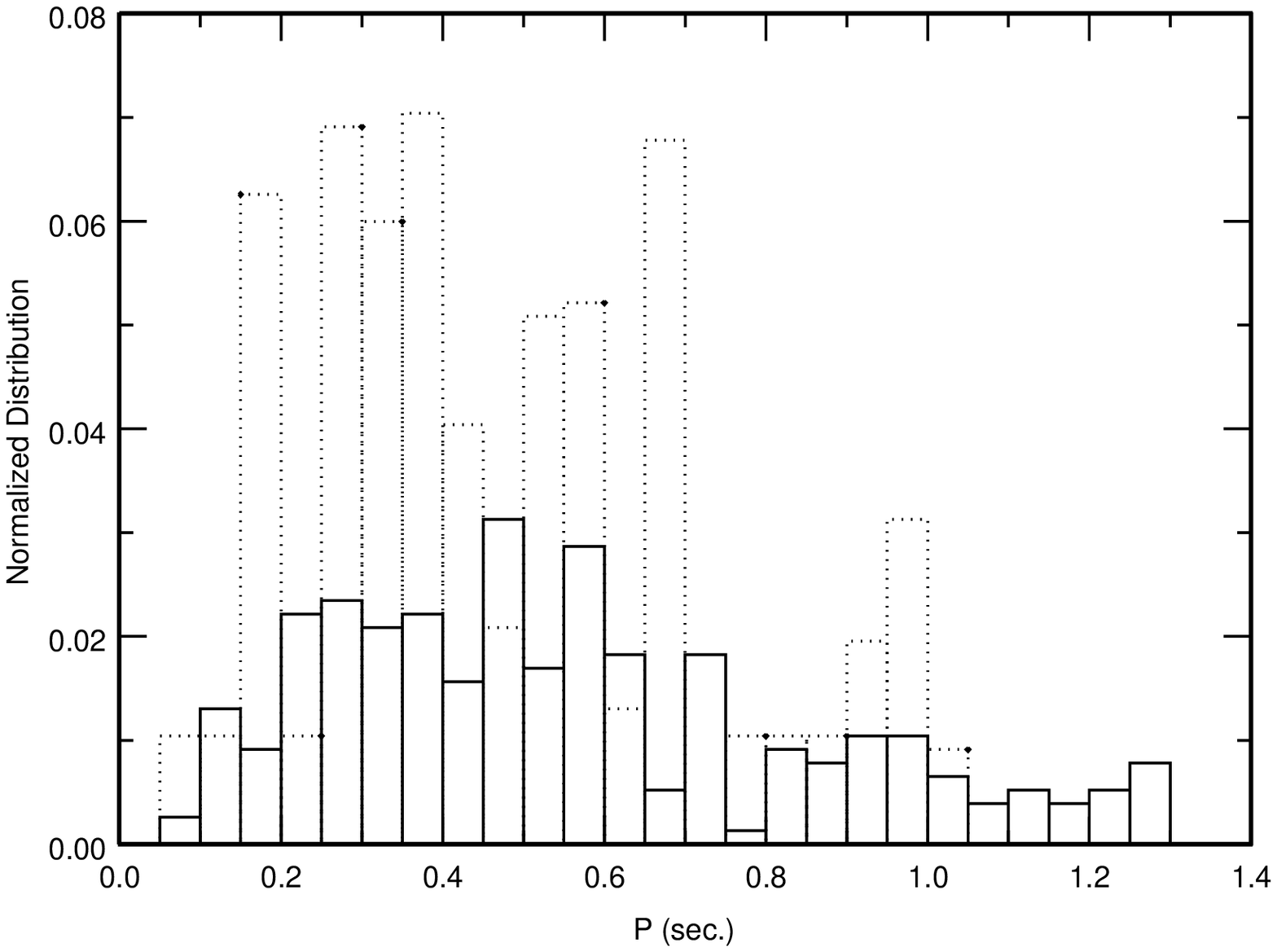}
\includegraphics{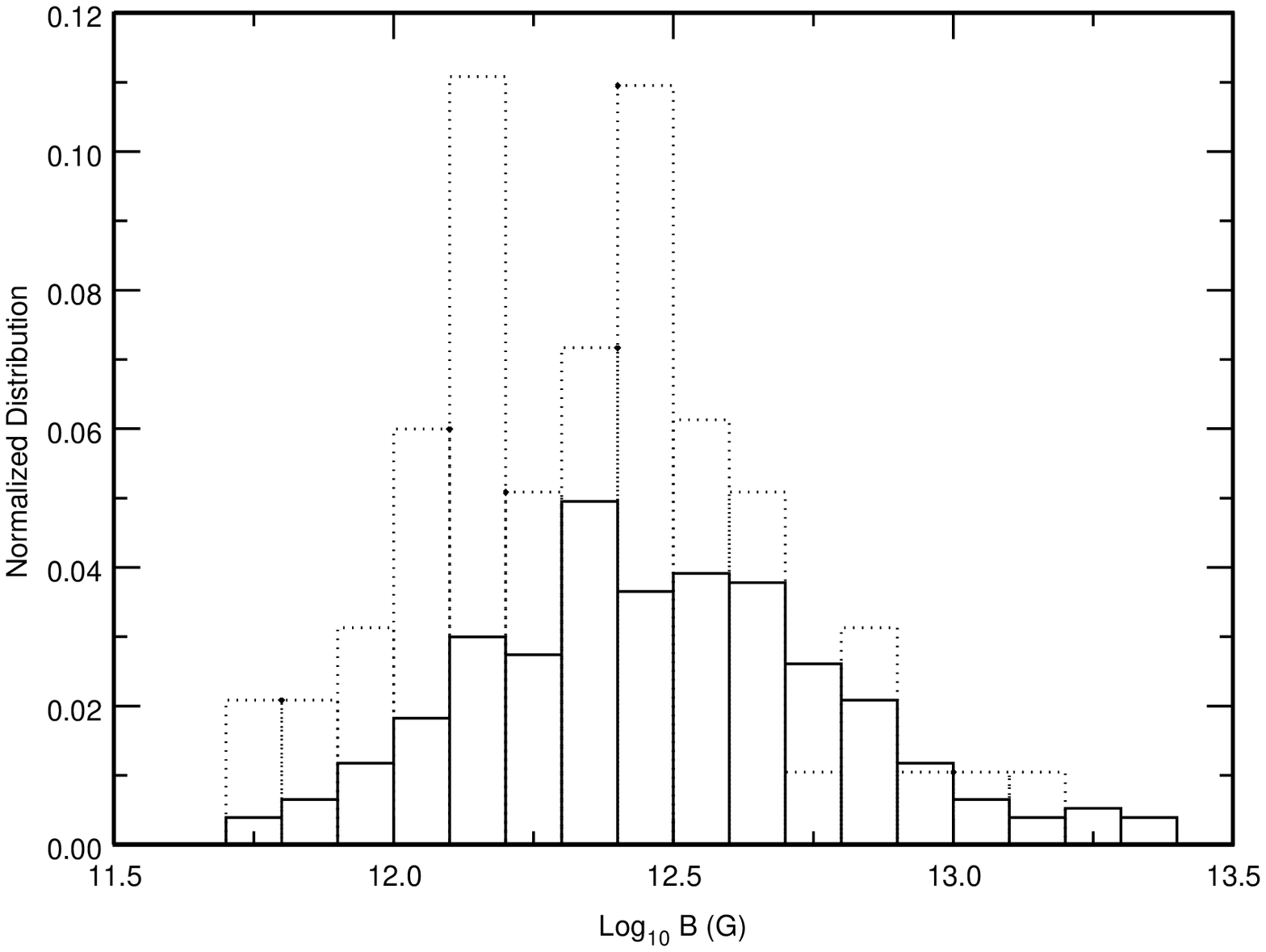} \includegraphics{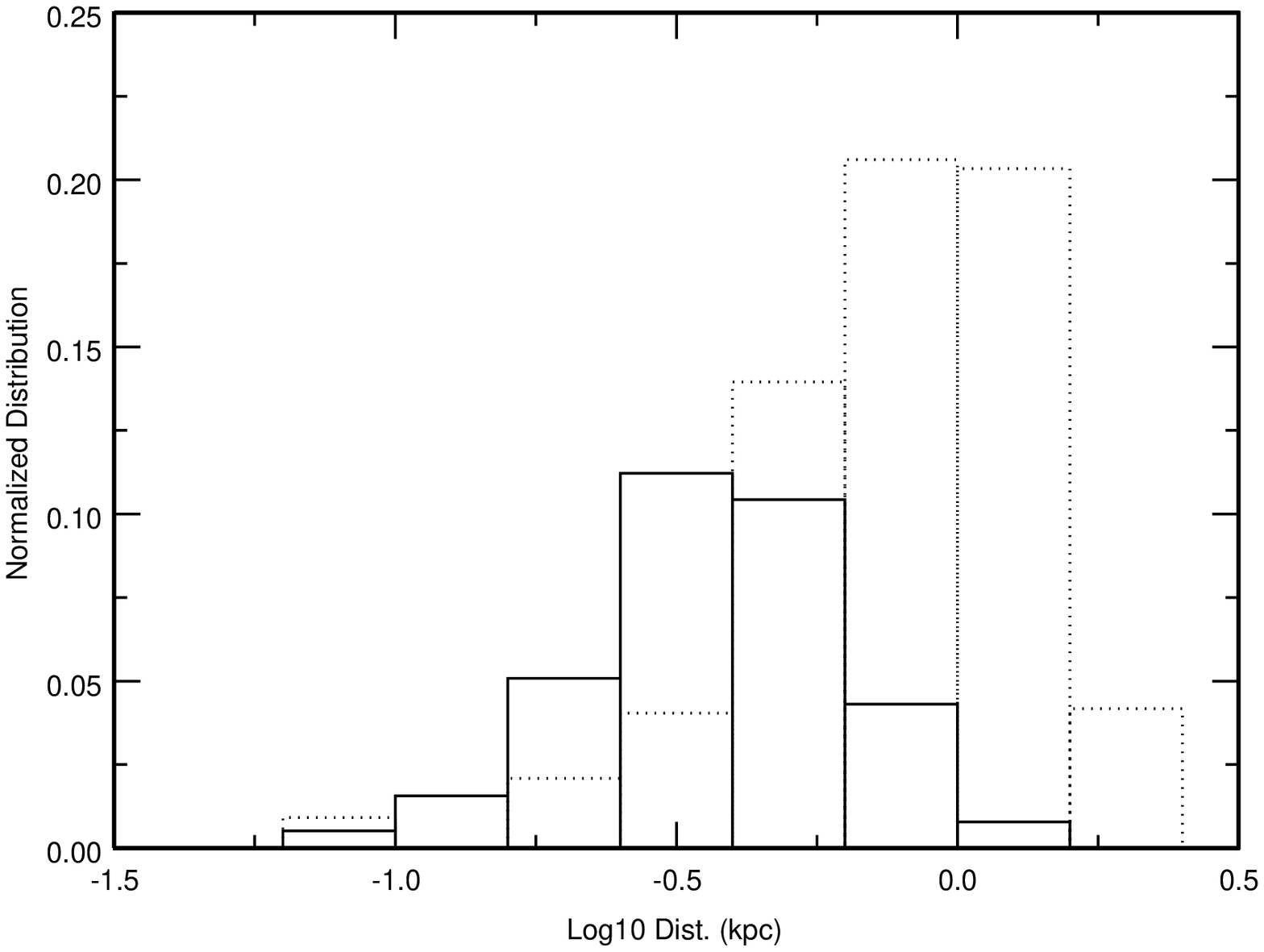} \includegraphics{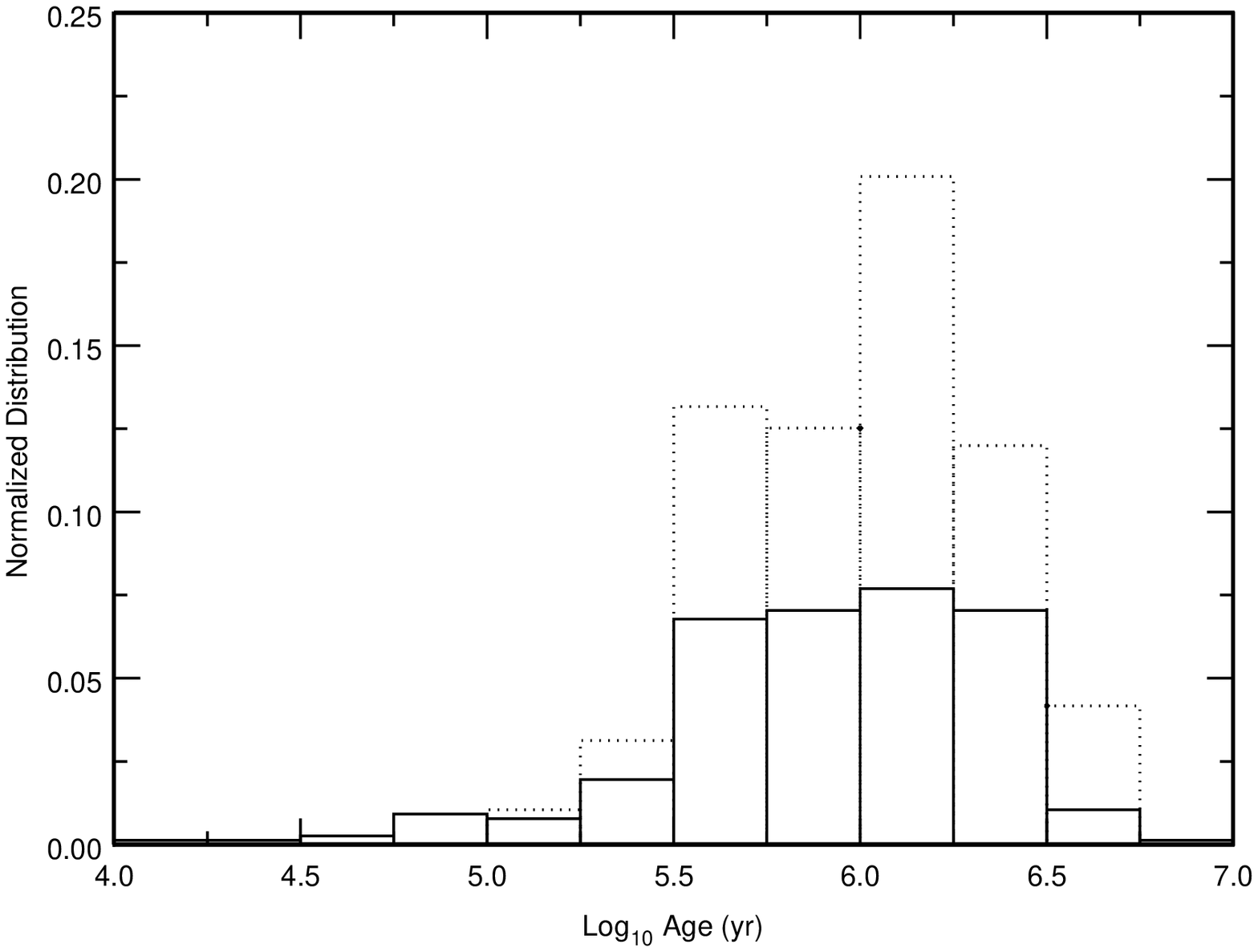}
 \noindent{Fig.~9 Normalized distributions of period,
 magnetic field, distance and age of the simulated $\gamma$-ray pulsars
 for $|b|>5^{\circ}$. Dotted histograms represent the distributions of
 the simulated $\gamma$-ray pulsars without those produced in the Gould belt, and solid
 histograms represent the simulated $\gamma$-ray pulsars only
 produced in the Gould belt.
 }
\end{figure}

\clearpage


\clearpage

\begin{thebibliography}{}
\singlespace
\bibitem[]{}
Arons, J. \& Scharlemann, E.T., 1979, \apj, 231, 854
\bibitem[]{618}
Bhattacharya, D., Wijers, R.A.M.J., Hartman, J. W., \& Verbunt,
F., 1992, A\&A, 254, 198
\bibitem[] {}
{\bf{Becker, W., et al. 2003, ApJ, 594, 798}}
\bibitem[]{621}
Biggs, J.D., 1990, \mnras, 245, 514
\bibitem[]{623}
Cheng, K. S., Ho, C. and Ruderman, M. A. 1986a, \apj, 300,500 (CHR
I)
\bibitem[]{626}
Cheng, K. S., Ho, C. and Ruderman, M. A. 1986b, \apj, 300,522 (CHR
II)
\bibitem[]{629}
Cheng, K. S. \& Zhang, J. L., 1996, ApJ, 463, 271
\bibitem[]{631}
Cheng, K. S. \& Zhang, L., 1998, \apj, 498, 327
\bibitem[] {}
{\bf{Cheng, K. S. \& Zhang, L., 1999, ApJ, 515, 337}}
\bibitem[]{633}
Cheng, K. S., Ruderman, M. \& Zhang, L., 2000, ApJ, 537, 964
\bibitem[]{635}
Daugherty, J.K. \& Harding, A. K. 1996, \apj, 458, 278
\bibitem[]{637}
Emmering, R.T. \& Chevalier, R. A. 1989, \apj, 345, 931
\bibitem[]{639}
Fan, G. L., Cheng, K. S. \& Manchester, R. N., 2001, ApJ, 557, 297
\bibitem[]{641}
Fichtel et al. 1994, \apj, 434, 557
\bibitem[]{643}
Gehrels, N., Macomb, D. J. \& Bertsch, D. L. et al. 2000, Nature,
404, 6776
\bibitem[]{646}
Gonthier, P. L., Ouellette, M. S., Berrier, J., O'Brien, S. \&
Harding, A. K., 2002, ApJ, 565, 482
\bibitem[]{649}
Grenier, I., 1997, Workshop on high energy cosmic neutrinos:
origin, production and detection. June 2-3, 1997 - Marseille,
France, "Gamma-ray sources and diffuse emission above 100 MeV"
\bibitem[]{653}
Grenier, I. A. , 2000, A\&A, 364, L93
\bibitem[]{655}
Grenier, I. A., 2003, in Texas in Tuscany: XXI Symposium on
Relativistic Astrophysics (Singapore: World Scientific), preprint
(astro-ph/0303498)
\bibitem[]{659}
Guillout, P., Sterzik, M. F., Schmitt, J. H. M. M. et al. 1998,
A\&A, 337, 113
\bibitem[] {}
{\bf{Haberl, F. et al. 2003, A\&A, 403, L19}}
\bibitem[]{662}
Halpern, J.P. \& Ruderman, M. A. 1993, \apj, 415, 286
\bibitem[]{664}
Halpern, J. P., Gotthelf, E. V., Mirabal, N. \& Camilo, F., 2002,
ApJ, 573, L41
\bibitem[]{667}
Harding, A. K. \& Muslimov, A. G. 1998, \apj, 508, 328
\bibitem[]{669}
Harding, A. K. \& Zhang, B., 2001, ApJ, 548, L37
\bibitem[]{671}
Hartman, R. C., Bertsch, D. L. Bloom, S. D. et al. 1999, ApJS,
123, 79
\bibitem[]{}
Hirotani, K. \& Shibata, S., 2001, \apj, 558, 216
\bibitem[]{674}
Kaaret, P. \& Cottam, J., 1996, \apj, 462, L35
\bibitem[]{}
Kramer, M. et al., 2003, \mnras, 342, 1299
\bibitem[]{676}
Lorimer, D. R., Bailes, M. \& Harrison, P. A., 1997, \mnras, 289,
592
\bibitem[]{679}
Manchester R. N., et al. 2001, MNRAS 328, 17
\bibitem[]{681}
McLaughlin, M. A., Mattox, J. R., Cordes, J. M. \& Thompson, D.
J., 1996, ApJ, 473, 763
\bibitem[]{684}
Mirabel, N., Halpern, J. P., Eracleous, M. et al. 2000, ApJ, 541,
180
\bibitem[]{687}
Mirabal, N. \& Halpern, J. P. , 2001, ApJ, 547, L137
\bibitem[]{689}
Montmerle, T., 1979, \apj, 231, 95
\bibitem[] {666}
{\bf{Narayan, R. \& Ostriker, J.P., 1990, ApJ, 352, 222}}
\bibitem[]{691}
Nolan, P. L., Tompkins, W. F., Grenier, I. A. \& Michelson, P. F.,
2003, ApJ accepted, preprint (astro-ph/0307188)
\bibitem[]{694}
Paczynski, B. 1990, \apj, 348, 485
\bibitem[Press et al. 1992]{press92}
Press, W., Flannery, B., Teukolsky, S., Vetterling, W. 1992,
Numerical Recipes: The Art of Scientific Computing 2nd ed.,
Cambridge Univ. Press, Cambridge
\bibitem[]{700}
Romero, G. E., Combi, J. A. \& Colomb, F. R., 1994, A\&A, 288, 731
\bibitem[]{702}
Romero, G. E., Benaglia, P. \& Torres, D. F., 1999, A\&A, 348, 868
\bibitem[] {}
{\bf{Sanwal, D., Pavlov, G., Zavlin, V. \& Teter, M. 2002, ApJ,
574, L61}}
\bibitem[]{704}
Sturner, S. J. \& Dermer, C. D. 1996, \apj, 461, 872
\bibitem[]{706}
Tavani, M., Barbiellini, G., Argan, A., et al. , 2001, Gamma 2001:
Gamma-Ray Astrophysics, Edited by Ritz, S., Gehrels, N. \&
Shrader, C. R., in AIP Conf. Proc., Vol. 587, p.729
\bibitem[]{710}
Torres, D. F., Butt, Y. M. \& Camilo, F., 2001, \apj, 560, L155
\bibitem[]{712}
Torres, D \& Nuza, S. E., 2003, \apj, 583, L25
\bibitem[]{714}
Yadigaroglu, I. A. \& Romani, R. W. 1995, \apj, 449, 211
\bibitem[]{716}
Yadigaroglu, I. A. \& Romani, R. W., 1997, \apj, 476, 347
\bibitem[]{718}
Zhang, L. \& Cheng, K. S. 1997, \apj, 487, 370
\bibitem[]{720}
Zhang, L. \& Cheng, K. S., 1999, \apj, 526, 327
\bibitem[]{722}
Zhang, L., Zhang, Y. J. \& Cheng, K. S. 2000, A\&A, 357, 957
{\bf{\bibitem[]{724} Zhang, L., Cheng, K.S., Jiang, Z. J. \&
Leung, P. 2004, \apj, in press (astro-ph0402089)}}
\end{thebibliography}
\end {document}